\documentclass[namedreferences]{SolarPhysics}
\usepackage{spr-sola-addons} 
\usepackage{color}           
\usepackage{url}             

\usepackage[pdftex]{graphicx} 

\newcommand{\degr}{{\hbox{$^\circ$}}}
\newcommand{\arcmin}{{\hbox{ $^\prime$}}}
\newcommand{\arcsec}{{\hbox{$^{\prime\prime}$}}}
\renewcommand{\sun}{_\odot}

\renewcommand{\tabcolsep}{0.18cm}

\newcommand{\be}{\begin{equation}}
\newcommand{\ee}{\end{equation}}
\newcommand{\bea}{\begin{eqnarray}}
\newcommand{\eea}{\end{eqnarray}}
\newcommand{\beas}{\begin{eqnarray*}}
\newcommand{\eeas}{\end{eqnarray*}}

\newcommand{\apj}{    {\it Astrophys. J.}}

\newcommand{\mnras}{  {\it Mon. Not. Roy. Astron. Soc.}}

\newcommand{\solphys}{{\it Solar Phys.}}

\chardef\us=`\_ 

\begin{document}

\begin{article}

\begin{opening}
\title{Modeling of the sunspot-associated microwave emission using a new method of DEM inversion\footnote{This work is dedicated to the memory of our good colleague and friend Anatoly Nikolaevich Korzhavin, who passed away on December 25, 2017.}}

\author[addressref={aff1},corref,email={calissan@cc.uoi.gr}]{\inits{C.E.}\fnm{C. E.}~\lnm{Alissandrakis}}
\author[addressref={aff2,aff3}]{\inits{V. M.}\fnm{V. M.}~\lnm{Bogod}}
\author[addressref={aff2}]{\inits{T. I.}\fnm{T. I.}~\lnm{Kaltman}}
\author[addressref={aff1}]{\inits{S.}\fnm{S.}~\lnm{Patsourakos}}
\author[addressref={aff2}]{\inits{N. G.}\fnm{N. G.}~\lnm{Peterova}}

\address[id=aff1]{Section of Astro-Geophysics, Department of Physics, University of Ioannina, GR-45110 Ioannina, Greece}
\address[id=aff2]{St. Petersburg branch of Special Astrophysical Observatory (Spb SAO), Russia} 
\address[id=aff3]{Saint-Petersburg National Research University ITMO, St. Petersburg, Russia} 

\runningauthor{C. E. Alissandrakis et al.}
\runningtitle{DEM inversion and sunspot modeling}

\begin{abstract}
We developed a method to compute the temperature and density structure along the line of sight by inversion of the differential emission measure (DEM), under the assumptions of stratification and hydrostatic equilibrium. We applied this method to the DEM obtained from AIA observations and used the results, together with potential extrapolations of the photosheric magnetic field, to compute the microwave emission of three sunspots, which we compared with observations from the RATAN-600 radio telescope and the Nobeyama Radioheliograph (NoRH). Our DEM based models reproduced very well the observations of the moderate-size spot on October 2011 and within 25\% the data of a similar sized spot on March 2016, but predicted too low values for the big spot of April 14, 2016. The latter was better fitted by a constant conductive flux atmospheric model which, however, could not reproduce the peak brightness temperature of $4.7\times10^6$\,K and the shape of the source at the NoRH frequency. We propose that these deviations could be due to low intensity non-thermal emission associated to a moving pore and to an opposite polarity light bridge. We also found that the double structure of the big spot at high RATAN-600 frequencies could be interpreted in terms of the variation of the angle between the magnetic field and the line of sight along the sunspot.
\end{abstract}
\keywords{Radio Emission,  Active Regions; Sunspots, Magnetic Fields; Active Regions, Structure; Active Regions, Models; Spectrum, Ultraviolet}
\end{opening}

\section{Introduction} \label{intro} 
It is well established that sunspot-associated microwave emission is primarily due to gyroresonance (gr) emission (\citealp{1962SvA.....6....3Z, 1962ApJ...136..975K}), with free-free (ff) emission playing a minor role. In the case of the gr process, radiation is emitted at thin layers above the sunspot where the magnetic field corresponds to a low harmonic of the gyrofrequency. The absorption coefficient is a complicated function of the magnetic field intensity, $B$, the angle between the magnetic field and the line of sight, $\vartheta$, the electron temperature, $T_e$ and the electron density, $N_e$.

An important aspect of gr emission is that the variation of the absorption coefficient with $\theta$ is very sharp, being zero for $\vartheta=0$ and increasing to large values as $\vartheta$ approaches 90\degr. As a consequence, when a sunspot is not far from the center of the disk, the brightness temperature of the associated source has a minimum near its center, where the magnetic field is almost parallel to the line of sight. Around 6\,cm wavelength this minimum is very narrow and unobservable in the extraordinary mode (e-mode) emission, but it is rather broad in ordinary mode (o-mode) emission. The result is that, even in the case of axial symmetry, the sunspot image in circular polarization (Stokes $V$) shows a ring structure which in total intensity (Stokes $I$) is less pronounced. This was observationally verified by the first high resolution observations and detailed models. 
If the physical conditions depart from plane parallel, additional structure is expected in cm-$\lambda$ sunspot images. For example, \cite{1982ApJ...253L..49A} reported a case where $T_b$ was significantly reduced at the center of a large spot both in $I$ and $V$; this could not be interpreted in terms of angle effects and was attributed to a decrease of the electron temperature above the sunspot umbra \citep{1984ApJ...277..865S}. 

Cases have been reported where the gr mechanism is not sufficient to account for the observed brightness temperature. One such case is the so called ``neutral line sources" (\citealp{1993A&A...270..509A, 2006PASJ...58...21U, 2012AstBu..67..425B, 2015SoPh..290...37B}) which, as their name implies, are occasionally observed above neutral lines of the magnetic field in the early stages of active region evolution. ``Peculiar sources'' have been reported, also near the neutral line but with very steep spectra in the 2-4\,cm range (\citealp{1986ApJ...301..460A,1990IAUS..142..483G,1993ApJ...419..398L}). Another case is the emission associated with a moving spot (\citealp{1987SoPh..112...89C,2008SoPh..249..315U}). In such cases, a small population of non-thermal electrons has been invoked for their interpretation (\citealp{1984A&A...131..103C,1993A&A...270..509A}).

After the first high resolution observations \citep{1977ApJ...213..278K}, detailed modeling using extrapolations of the photospheric magnetic field and simple atmospheric models succeeded in reproducing both the observed brightness temperatures and the structure of the microwave sunspot associated sources \citep{1980AA....82...30A}. Other similar computations relied on models of the sunspot field rather than extrapolations of the observed (\citealp{1979AZh....56..562G,1985A&A...143...72K}). Using RATAN-600 data and a dipole magnetic field \cite{2010AstBu..65...60K} obtained  temperature-height profiles for sources located above sunspots. More recently a diagnostic technique was developed by the Pulkovo group, based on RATAN-600 observations and photospheric magnetic field extrapolation, that makes possible to estimate the electron density and temperature of the emitting region  (\citealp{2012ARep...56..790K, 2013Ge&Ae..53.1030K, 2015Ge&Ae..55.1124K,2018SoPh..293...13S}). Extensive modeling work has also been done by the group of the New Jersey Institute of Technology (\citealp{2015ApJ...805...93W, 2018ApJ...853...66N}), in anticipation of the Expanded Owens Valley Solar Array.

Radiation in extreme ultraviolet (EUV) lines is formed in the Transition Region (TR) and low corona, hence is ideal in providing information complementary to the microwave data. Although the EUV emission is not sensitive to the magnetic field, it is an important diagnostic of electron temperature and density, albeit a complicated one due to NLTE effects, uncertainties in elemental abundances and its optically-thin character. \cite{1996SoPh..166...55N} were the first to incorporate pressure measurements from EUV data in modeling the active region emission obtained in joint 1D spectral observations from RATAN-600 (0.8-31.6\,cm) and 2D images from the Very Large Array (VLA) at 6 and 20\,cm. They found that the pressure was higher by a factor of 1.54 above the plage, compared to the sunspot and that the effect of horizontal pressure variations was small at 6\,cm and more important at 20\,cm.  Subsequently, \cite{2000ApJS..130..485N}, using soft X-ray data, found that both the temperature and the emission measure (EM) were lower above sunspot umbrae than above penumbrae and, more recently, \cite{2011ApJ...728....1T} used qualitative EUV information from the Solar and Heliospheric Observatory (SoHO) in their analysis of combined microwave observations from the VLA and the Owens Valley Solar Array. In our days, the Atmospheric Imaging Assembly (AIA) instrument aboard the Solar Dynamics Observatory (SDO) is a very good source of EUV data. Algorithms have been developed for computing the differential emission measure (DEM) (e.g. \citealp{2012A&A...539A.146H, 2013ApJ...771....2P}), but this information has not been used so far in conjunction with microwave data.

In this article we present a method to derive electron density and temperature profiles from the DEM computed from AIA data and we use the results to model 1D spectral observations from RATAN-600 and 2D images from the Nobeyama Radioheliograph (NoRH), while retaining the plane-parallel approximation. Our sunspot sample includes two moderate-size sunspots and a large sunspot that showed spatial structure in  RATAN-600 and NoRH observations, in spite of the moderate instrumental resolution. In Section 2 we describe the observations, in Section 3 we present the DEM analysis of a sunspot, in Section 4 we describe our method of computing $T_e$ and $N_e$ as a function of height from the DEM and in Section 5 we give our results of modeling the microwave emission from the two moderate-size spots and for the large sunspot observed on April 14, 2016. Finally, in Section 6 we summarize our results and present our conclusions.

\section{Observations and data reduction}

\subsection{Microwave observations and processing}\label{radobser}
The NoRH is a T-shaped synthesis instrument that observes routinely the sun at 17 and 34\,GHz with a resolution of $\sim15$\arcsec\ at the lower frequency. The NoRH 17\,GHz I and V images that we used were obtained near local noon, at 02:45 UT, and were retrieved from the instrument's ftp archive; no sunspot-associated emission was detected at 34\,GHz. 

RATAN-600 has a knife-edge beam and provides one-dimensional (1D) images in the cm-$\lambda$ range; at 2 cm the beam size is  17\arcsec\ by 15\arcmin. Observations are made using the southern sector of the 600-m diameter ring and the periscopic mirror (\opencite{2011AstBu..66..190B}). The radio telescope has a large effective area (300-600\,m$^2$) in the centimeter range. It provides a flux sensitivity of $\sim0.01$\,sfu, with position accuracy of about 1\arcsec. It covers a broad frequency band (3-18\,GHz) with high spectral resolution (0.15\,GHz) and provides high-precision measurement of the polarized signal, 0.05 to 0.2\%. The digital multi-channel data recording system (\opencite{2011AstBu..66..205B}) provides a large dynamic range, which extends from the instrumental noise level (antenna temperature, $T_A$, of about 300 to 500\,K) up to high-power signals during flares ($T_A\sim200$ to 500$\times10^3$\,K). This corresponds to about 30 to 80 times the quiet Sun level.

Before measuring the source parameters from the RATAN-600 1D scans, we need to determine the quiet sun background. We did this by fitting the scans up to $\pm900$\arcsec\ from the disk center to a parabolic curve of the form $T_A(x)=x_0-bx^2$ with $b>0$, excluding the regions of strong emission. Subsequently the flux, the peak $T_A$, the position and the width of the source were measured; we converted the flux and $T_A$ to absolute units by using the standard calibration factor for the instrument.  We note that the background level affects less the peak antenna temperature than it affects the flux; the peak $T_A$ is also less affected by weak non-spot emissions that might exist near the sunspot source.

\subsection{EUV observations and processing}
Our EUV data are from the AIA instrument aboard SDO, which  operates continuously and provides high cadence and high resolution images in six spectral bands around 171, 193, 211, 131, 335 and 94\,\AA, covering a range of $\log T_e$ from $\sim5.5$ to $\sim7.5$.

We computed the DEM from AIA images using the algorithm of  \cite{2013ApJ...771....2P}. This method uses the observed AIA intensity in its 6 coronal channels; it is a very fast iterative method, based on regularization, where the temperature response functions of the AIA channels are used as basis functions in the inversion.  We note that it is well-known that the DEM inversion problem is an ill-posed one, i.e., there are more unknowns than knowns, and thus the computed solutions are not unique. Moreover, the method occasionally gives negative values for the DEM; these were ignored in our further analysis.

\begin{figure}[h]
\begin{center}
\includegraphics[height=4.2cm]{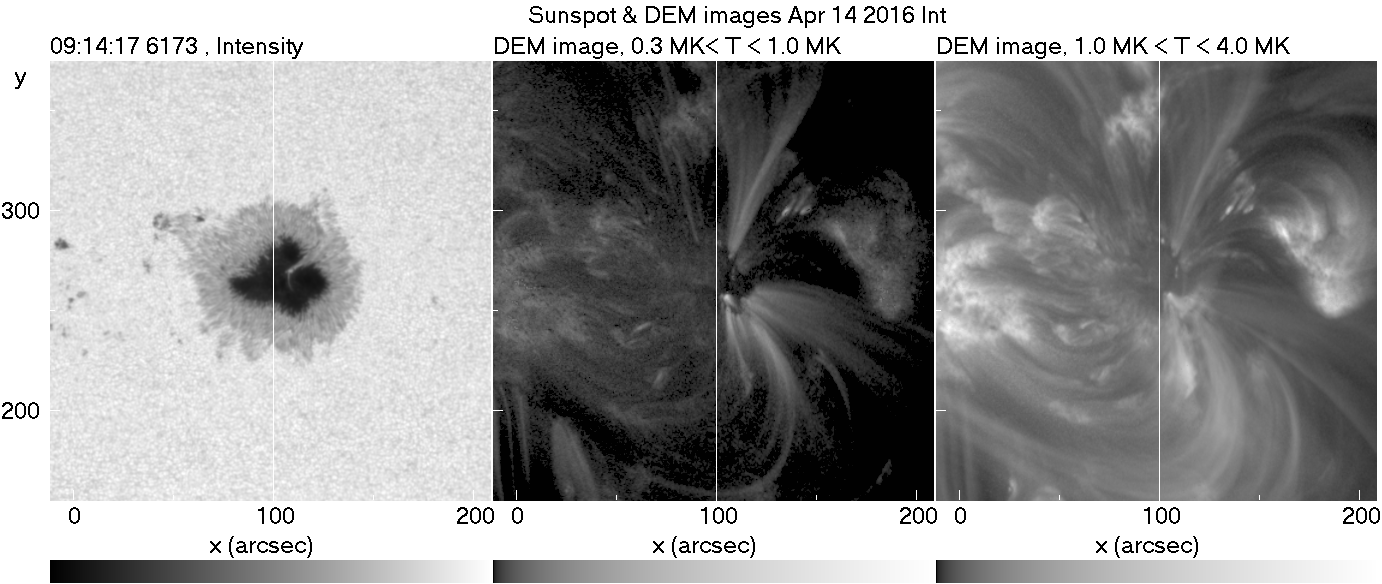}\includegraphics[height=4.2cm]{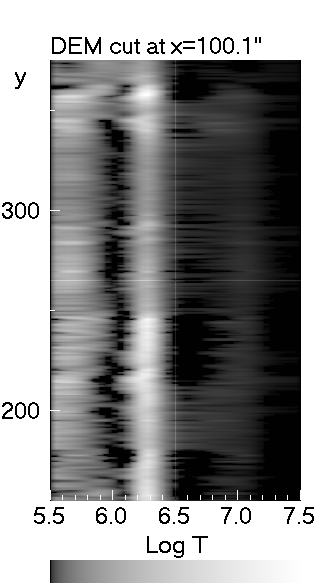}
\end{center}
\caption{Images of the sunspot of April 14, 2016 and of the computed DEM in two temperature ranges. DEM images are displayed with the same range of values ($1.3\times10^{26}$ to $8.1\times10^{27}$\,cm$^{-5}$) and the same contrast. At the right a cut of DEM, as a function of $\log T_e$ and position along the white line of the images at the left; values are in the  range of 0 to $3.46\times10^{21}$\,cm$^{-5}$K$^{-1}$. The orientation in this figure is with the solar north up.}
\label{dem}
\end{figure}

\section{Differential emission measure from AIA}\label{DEM_AIA}
Figure \ref{dem} gives the results of our DEM computations for the large sunspot of April 14, 2016, which will be further discussed in section \ref{Apr16}. The N-S DEM cut through the sunspot (right) shows three maxima in $T_e$: a strong maximum around $\log T_e=6.27~(T_e=1.9\times10^6$\,K), a more weak one at $\log T_e=5.65~(T_e=4.47\times10^5$\,K) and a much weaker at $\log T_e=6.98~(T_e=9.12\times10^6$\,K). Images of the DEM in the first two temperature ranges are given in Figure \ref{dem}. The high-temperature peak can be rather safely dismissed because (a) it is unlikely to have such high temperatures outside flares, particularly in sunspots and quiet-Sun regions, (b) the peak value, log(DEM) $\sim18.5$\,cm$^{-5}$ yields unrealistically small densities of $\sim2.5\times{10}^4$\,cm$^{-3}$ for a standard coronal scale height of 50\,Mm and (c) this DEM peak is very weak, typically 0.01 of the corresponding primary peaks. The required densities could be even smaller, if we consider the scale-height corresponding to $T_e$, which would be $\approx \sqrt{9.5}$ larger than the scale-height for 1\,MK (i.e., $\approx$ 50 Mm). One may attribute such weak secondary peaks to limitations inherent to DEM inversions of narrow-band data (e.g., limited number of channels, non-uniform temperature coverage, secondary peaks of the temperature response functions). We note that active region cores, particularly when undergoing flaring, do show evidence of 10\,MK plasmas (e.g. \citealp{2013PASJ...65S...8A, 2016A&A...588A..16S}).

\begin{figure}[h]
\begin{center}
\includegraphics[width=.45\textwidth]{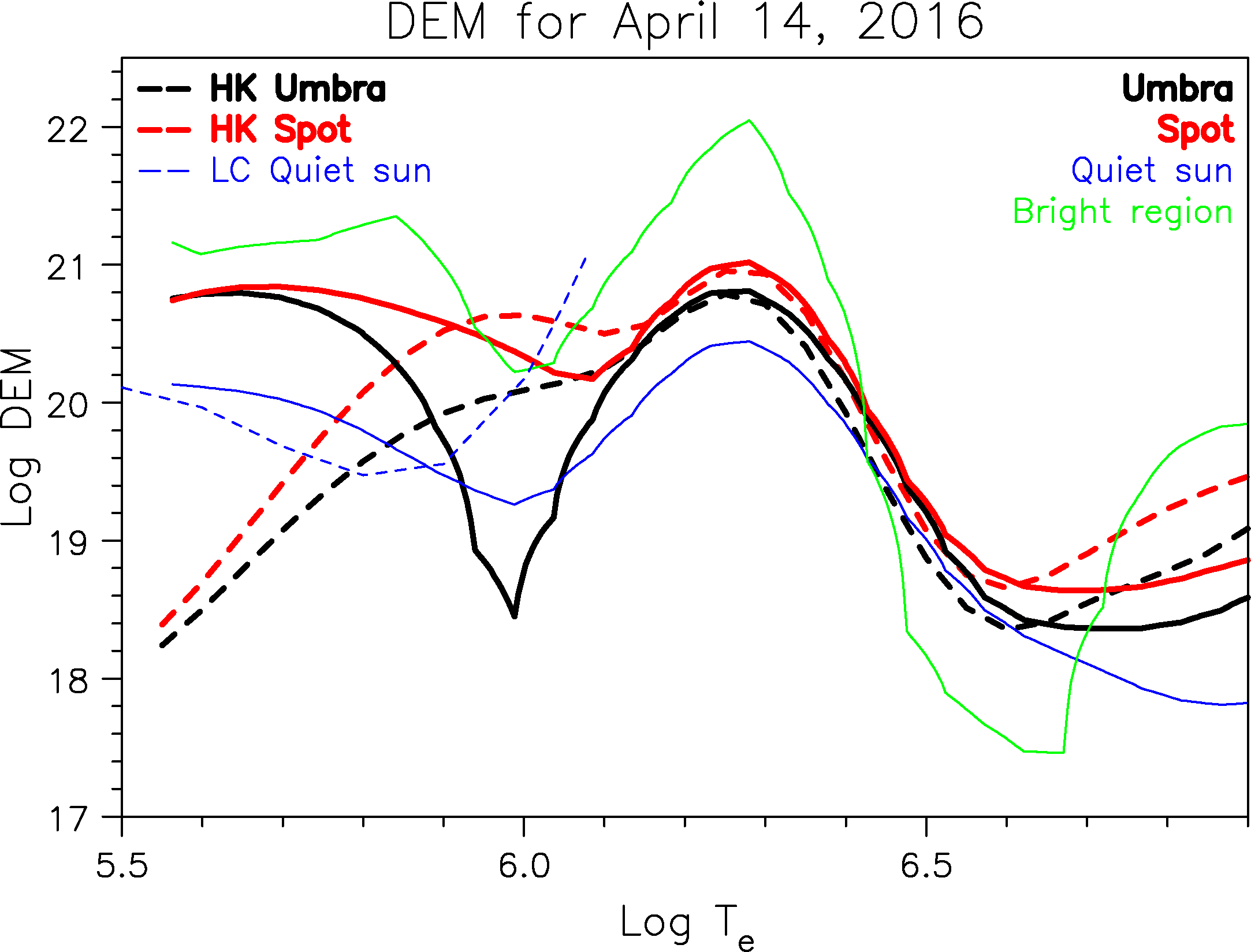}
\end{center}
\caption{The average DEM over the umbra, the entire spot, a quiet sun region and a bright region. Full lines show the results from the method of \cite{2013ApJ...771....2P}; dashed lines labeled HK show the results from the method of \cite{2012A&A...539A.146H} and the dashed line labeled LC gives the DEM from \cite{2008ApJ...675.1629L}. The DEM is in cm$^{-5}$K$^{-1}$.}
\label{EM0}
\end{figure}

Plots of the DEM as a function of $T_e$ for $\log T_e < 6.9$, averaged over an 8 by 8\arcsec\ region of the umbra, together with the DEM averaged over the entire spot (dominated by the penumbral contribution) are given in Figure \ref{EM0}. DEM curves for a very quiet 100 by 100\arcsec\ region in the NW of our field of view and a 9 by 9\arcsec\ bright region at $x=-3.4,~y=241.8$\arcsec\  are also given for reference. Also for reference we give the DEM for the quiet sun derived from combined UV and radio observations by \cite{2008ApJ...675.1629L}.

A first remark is that the umbra and sunspot DEM is above that of the quiet sun region and below that of the bright region. The peak values in log DEM for the full spot and the umbra are 21.02 and 20.81 respectively, differing by a factor of 1.6 and the EM over the available $T_e$ range is $6.39\times10^{27}$\,cm$^{-5}$ and $4.01\times10^{27}$ respectively, with the same ratio as the peak DEM; this implies a density ratio of 1.3. For comparison, \cite{2009A&A...505..307T}, using SUMER spectra, give similar peak log DEM values, 21.5 and 21.2 for the penumbra and the umbra respectively, but their peaks are at lower $\log T_e$ with respect to ours, 5.7 and 5.75 respectively. The lower temperature DEM sunspot peaks of \cite{2009A&A...505..307T} might be attributed to the fact that  formation temperatures of the SUMER spectral lines used in that study did not exceed $\sim1$\,MK, thus they were not able to probe hotter plasmas. On the other hand, while AIA has a better coverage of the coronal part of the plasma thermal distribution, it nevertheless lacks the transition region coverage of SUMER.  Finally, both \cite{2009A&A...505..307T} and our study employ different sunspots. It is obvious that such studies should be undertaken over larger sunspot samples. We add that the recent sunspot model of \cite{2015ApJ...811...87A} predicts that the DEM maximum is at still lower temperature, at $\log T_e=5.5$ with a value of 20.5 in log DEM. As for the quiet sun DEM of \cite{2008ApJ...675.1629L}, it is very close to our DEM up to $\log T_e\sim5.9$.

In order to check our results, we also performed calculations using the algorithm of \cite{2012A&A...539A.146H}. The results, shown by dashed lines in Figure \ref{EM0}, are very similar to those from the method of \citealp{2013ApJ...771....2P} for the main peak, while the low temperature peak is shifted to higher values and the region above $\log T_e=6.6$ is enhanced. Thus the computation appears robust, at least as far as these two methods and the main peak are concerned. 

\section{DEM inversion}\label{invertDEM}
In the case of optically thin free-free emission, the optical depth and the brightness temperature in the radio range can be computed directly from the DEM, as was done by \cite{2013PASJ...65S...8A}. However, sunspot emission is optically thick and hence, in order to compute the microwave emission, we need information about the variation of $N_e$ and $T_e$ with height. In order to obtain this information from the DEM, additional assumptions are required.

To this end, \citet{1975SoPh...45..301W} used EUV spectral-line intensities recorded over quiet and active regions to infer their DEM. He then proposed a set of schemes based on either plane-parallel atmospheric models in hydrostatic equilibrium and  constant pressure with the temperature gradient controlled by thermal conduction or cylindrical models pertinent to spicules, prominence threads and coronal loop sections to infer the variation of temperature and density with height from observed DEM. This methodology was applied to an active region loop system by \citet{1977SoPh...51...83L}, who used observations of emissions measures, along with the assumption of hydrostatic equilibrium and constant thermal conduction flux, to derive the variation of pressure with temperature. \citet{1978ApJ...220..643R} gave  an analytical expression of the DEM for static, isobaric and steady coronal loops, that could be used to infer the variation of temperature along the loops. \citet{1992MNRAS.256...37H} described empirical models of the TR of bright giants, in which he recovered the variation of temperature with distance from the DEM, under the assumption of hydrostatic equilibrium and taking into account non-thermal motions.   

Our analysis starts with the necessary assumption that both  $N_e$ and $T_e$ vary monotonically along the line of sight, which is a reasonable one for the sunspot umbra but not for structures such as plumes, where dense/hot material is between regions of less dense and probably cooler plasma. A plausible additional assumption is hydrostatic equilibrium, which should be valid unless we have high velocity motions capable of destroying it. Under this assumption we have:
\be
\frac{dP}{dz} = -g \rho    \label{hydro}
\ee
where $P$ is the gas pressure, $\rho$ the density and $g$ the gravity. Expressing $P$ and $\rho$ in terms of $N_e$, (\ref{hydro}) gives:
\be
dz=-\frac{k}{\mu g  m_{\rm{H}}}\frac{d(N_{\rm{e}}T_e)}{N_e} \label{hydro2}
\ee
where $\mu$ is the mean molecular weight and $m_{\rm{H}}$ the mass of the hydrogen atom. Using
\be
p=N_eT_e
\ee
instead of the pressure, (\ref{hydro2}) gives:
\be
dz=-\frac{kT_e}{\mu g  m_{\rm{H}}}\frac{dp}{p} \label{eq16}
\ee
and, from the definition of the DEM, $\varphi$, we obtain:
\bea
\varphi(T_e)=N_e^2\frac{d\ell}{dT_e}&=&\frac{1}{\cos\alpha}\frac{p^2}{T_e^2}\frac{dz}{dT_e} \label{eq5} \\
                                                        &=&-\frac{k}{\mu g \cos\alpha\, m_{\rm{H}}} \frac{p}{T_e} \frac{dp}{dT_e}
\eea
where $\ell=z/\cos\alpha$ is the distance along the line of sight, with $\alpha$ being the heliocentric angle of the region. The above equation gives:
\be
p\,dp=-\frac{\mu\, g \cos\alpha\, m_{\rm{H}}}{k} \varphi(T_e)\, T_e\, dT_e
\ee
Assuming constant $\mu$ and $g$, this integrates to:
\be
p^2(T_{e})=p^2(T_{e_1})-2\frac{\mu\, g \cos\alpha\, m_{\rm{H}}}{k}\int_{T_{e_1}}^{T_{e}} \varphi(T_e)\, T_e\, dT_e  
\label{eq21}
\ee
where the integration constant, $p(T_{e1})$, is the value of $p$ at $T_e=T_{e_1}$. This expression is similar to equation (9) of \citet{1992MNRAS.256...37H}. We note that in order to keep the rhs of (\ref{eq21}) positive, the integration constant must satisfy the condition:
\be
p^2(T_{e1}) > 2\frac{\mu\, g\,\cos\alpha\  m_{\rm{H}}}{k}\int_{T_{e1}}^{T_{e2}} \varphi(T_e)\, T_e\, dT_e \label{pmin}
\ee
which provides a lower limit to $p(T_{e1})$, depending on the chosen values of $T_{e1}$ and $T_{e2}$.

Equation (\ref{eq21}) provides the variation of $p$ with $T_e$; once this is known, the corresponding height can be obtained from (\ref{eq5}):
\be
z(T_{e2})=z(T_{e1})+\cos\alpha\, \int_{T_{e1}}^{T_{e2}} \varphi(T_e)\, \frac{T_e^2}{p^2}\, dT_e \label{height}
\ee
where the second integration constant, $z(T_{e_1})$, is the height at the location where the temperature is equal to $T_{e_1}$. In practice it is convenient to specify the first integration constant at the highest temperature and the second at the lowest (i.e. at the base of the TR). We tested the numerical integration for a TR specified by constant conductive flux and hydrostatic equilibrium and found residuals smaller than 1\%. 

In the above both $\mu$ and $g$ were assumed constant. For fully ionized H and He, $\mu=0.61$; we obtained the same value from model C7 of \cite{2008ApJS..175..229A}, practically constant for $T_e>10^5$\,K. As for $g$, which varies $\propto{(R_{\sun} +z)})^{-2}$, the variation is 9\% for $z=40$\,Mm and 30\% for $z=100$\,Mm. We note that (\ref{height}) does not contain $g$, whereas the integrations in both (\ref{height}) and (\ref{eq21}) are stepwise.
Thus we can take into account the variation of $g$ by first computing $z$ at each step from (\ref{height}), and then use the proper value of $g$ in (\ref{eq21}); in this case the integration constants must both be specified at the lowest temperature.

\begin{figure}[h]
\centering
\includegraphics[width=\textwidth,height=5.5cm]{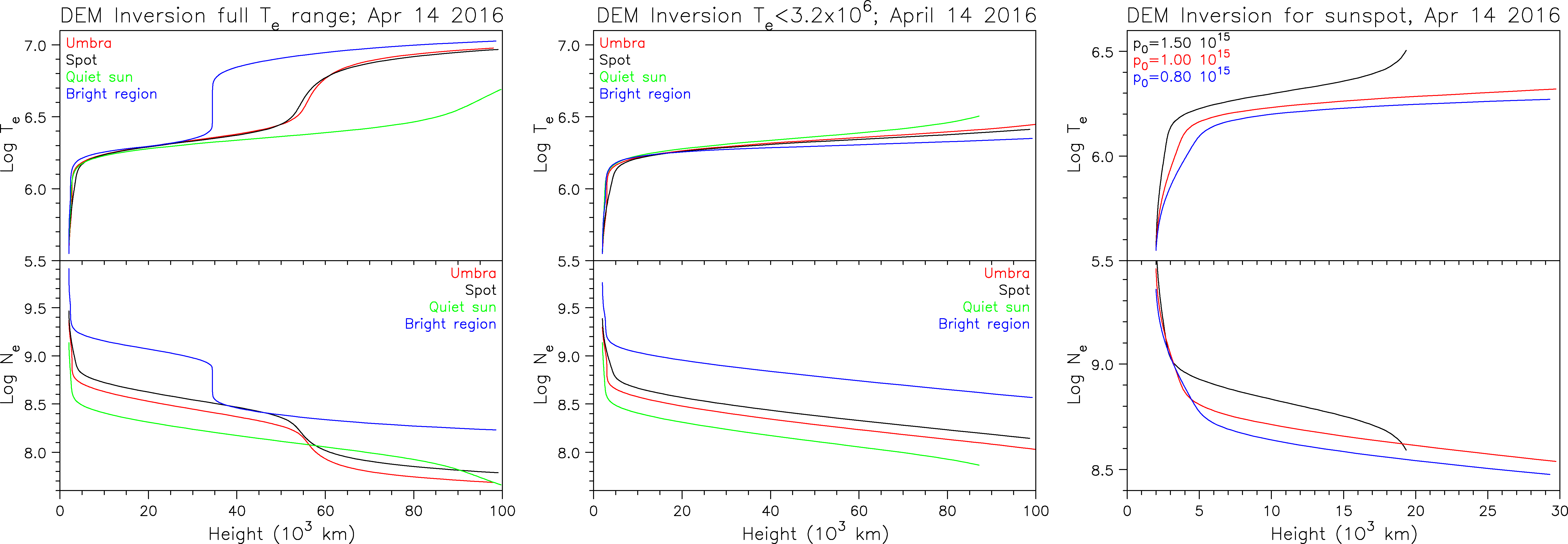}
\caption{Results of DEM inversion for April 14, 2016. Left: $T_e(z)$ and  $N_e(z)$ using the full DEM temperature range. Center: Inversion using only the DEM below the high temperature peak at $\log T\simeq7$. Right: Inversion for the sunspot DEM, for three values of the pressure integration constant; scales are different.}
\label{DEMinvert}
\end{figure}

The results of the inversion for the DEMs given in Figure \ref{EM0} are shown in the left panel of Figure \ref{DEMinvert}. The full temperature range was used, constant $g$ was assumed and the boundary values were set to $z(T_{e_{min}})=2000$\,km and $p(T_{e_{max}})=10^{13}$\,K\,cm$^{-3}$ which, for $T_{e_{max}}=2.8\times10^7$\,K, corresponds to $N_e=3.6\times10^5$\,cm$^{-3}$; we note that the value of $p(T_{e_{max}})$ does not affect much the structure of the lower layers. 

The inversion reproduces the sharp temperature rise in the TR, which is followed by a {\it plateau} at coronal temperatures. Higher up and beyond the region of formation of the gr emission, we have a second sharp rise and a second $T_e$ plateau, due to the high temperature peak in the DEM. This plateau disappears when we use only DEM values below the high temperature peak (central panel in Figure \ref{DEMinvert}), where we set $p(T_{e_{max}})=2\times10^{14}$\,K\,cm$^{-3}$ ($T_{e_{max}}=3.2\times10^6$\,K, $N_e=6.25\times10^7$\,cm$^{-3}$). Allowing for variable $g$ while keeping the same density values at the base of the TR, shifts the second sharp rise to a slightly larger height, without affecting much the region of gr emission.

The right panel of Figure \ref{DEMinvert} shows the dependence of the sunspot DEM inversion on the pressure integration constant, $p^2(T_{e_1})$, with the height range limited to the region of gr emission. For this case equation (\ref{pmin}) gives $p(T_{e1,min})=0.84\times 10^{15}$\,K\,cm$^{-3}$ and the figure shows plots for  $p(T_{e1})$ of 1.5, 1.0, and $0.8\times 10^{15}$; for the last value the maximum $T_e$ was limited to $2.39\times10^{6}$\,K. As $p(T_{e1})$ increases, $p$ increases and so does the temperature gradient, from (\ref{eq5}).  For very high values of $p(T_{e1})$ the plateau at coronal temperatures shrinks and this provides an empirical upper limit for $p(T_{e1})$.

\section{Modeling of the radio data}
\subsection{Models of the radio emission}\label{radiomodels}
In order to compute  the gr emission we need the magnetic field as a function of height, as well as a model of $T_e$ and $N_e$. The magnetic field above the photosphere can be derived by a potential or force-free extrapolation of the photospheric magnetic field (\citealp{1981A&A...100..197A}). For the purpose of this work we used a potential extrapolation since, as shown by \cite{1984A&A...139..271A} and \cite{1996SoPh..166...55N}, the introduction of currents in the linear force-free approximation just rotates the sunspot image, without affecting much the brightness temperature or the flux.

We used a plane-parallel approximation for $T_e$ and $N_e$, justified by the fact that our DEM computations showed a rather small density difference between the umbra and the penumbra. Two models for the variation of $T_e$ and $N_e$ with height were used: One (Model 1) was the model employed by \cite{1980AA....82...30A}, in which  the temperature variation with height in the transition region is specified by the condition of constant conductive flux and the $N_e$ profile by hydrostatic equilibrium. For a given magnetic field structure, this model is fully specified by three parameters: the height, $H_0$, at the base of the TR (where $T_e=10^5$\,K), the value of $p$ at that height, $p_0$, and the value of the conductive flux, $ F_c$. In the second model (Model 2) we used the temperature/density structure computed from the inversion of the DEM; below $\log T=5.5$, the model was joint to a constant $F_c$ model, with the value of $F_c$ fixed by the $N_e$ and $T_e$ values just above $\log T=5.5$.  Model 2 has only two free parameters, $H_0$ and $p_0$.

For all models we computed 2-D images of the emission over the sunspot region as a function of frequency, in R and L  polarizations, as well as in Stokes parameters I and V; the quiet sun background was subtracted and the flux was computed. The computed images were then integrated in the direction perpendicular to that of the RATAN-600 scans, corrected for instrumental resolution and converted to antenna temperatures; from the integrated scans we computed the peak $T_A$, the source position and the source width. 

In the case of the constant $F_c$ Model 1, which is parameterized, a grid of $\sim200$ models was computed, covering a range of values in $H_0$, $p_0$ and $F_c$. 
For the DEM Model 2 we had to use the highest value possible for $p_0$ (see last paragraph of section \ref{invertDEM}) in order to approach the observations and the lowest value of $H_0$ deduced from models 1.

The best model was selected using two procedures: (a) a least square fit of the model peak $T_A$ to the observed for all observed frequencies; we used the antenna temperatures rather than the flux for the reasons discussed in Section \ref{radobser} and (b) a least square fit of full 1D scans to the observed; in the second method we took into account the position shift between observed and model scans. The advantage of the second method is that it provides a global comparison of the model to the observations, which takes into account the width as well as the peak of the source, whereas the first is more efficient in isolating the sunspot-associated emission from the emission above the plage. The fit gave slightly different parameters for R and L, thus we also computed a combined fit for both polarizations; for this purpose, $\chi^2$ was normalized by the maximum observed $T_A(f)$ for fit procedure (a) and $T_A(x,f)$ for fit procedure (b), where $f$ is the frequency and $x$ the 1D position:
\be
\chi^2=\frac{1}{T_{A,max}}\sum (T_{A,obs}-T_{A,mod})^2
\ee
where the summation is over frequency for procedure (a) and over both frequency and position for procedure (b). The combined $\chi^2$ for R an L is:
\be
\chi^2_{RL}=\frac{1}{2}(\chi^2_{R}+\chi^2_{L})
\ee

\subsection{Observations and modeling of moderate-size sunspots}\label{small_spots}
\begin{figure}[ht]
\begin{center}
\includegraphics[width=0.42\textwidth]{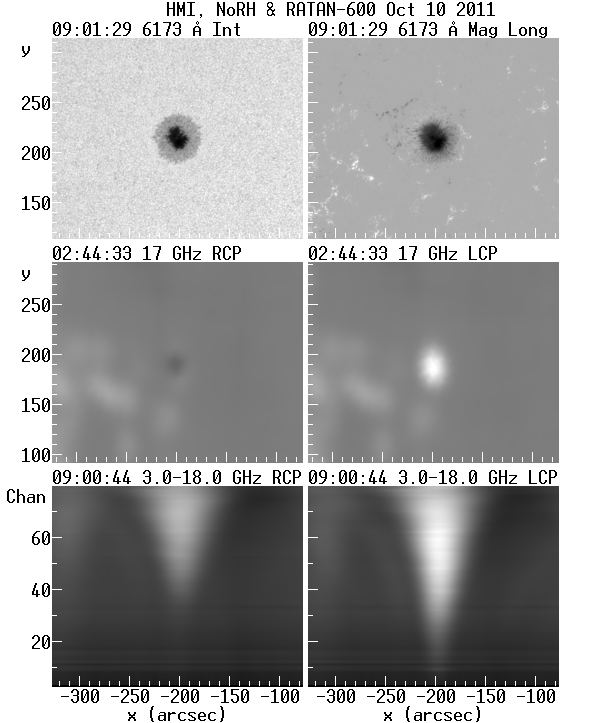}%
\includegraphics[width=0.42\textwidth]{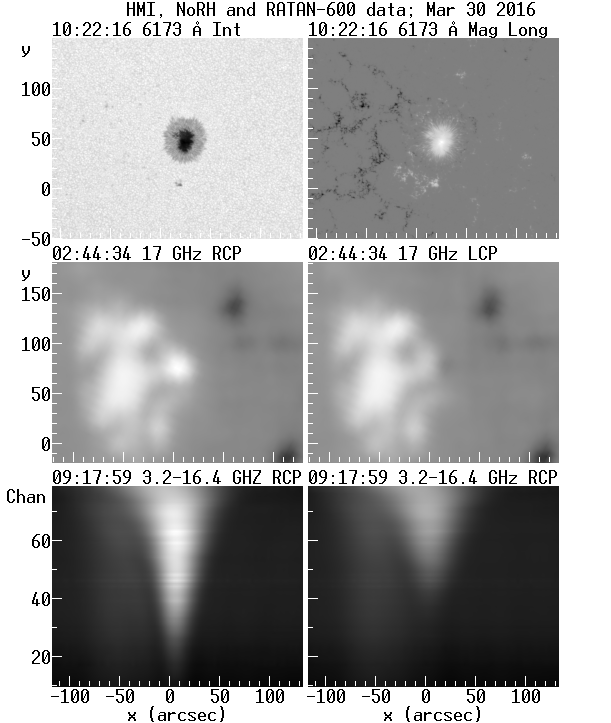}
\end{center}
\caption{Observations of sunspots in active regions 11312 (left) and 12526 (right), oriented with the celestial North up. The continuum intensity and the magnetic field are shown in the top row, the NoRH images in Right and Left circular polarization in the middle row and the RATAN-600 observations in R and L in the bottom row; the spectral range is from 3 to 18.0\,GHz for October 10, 2011 and from 3.18 to 16.4\,GHz for March 30, 2016.}
\label{otherspots}
\end{figure}

We start the presentation of the microwave emission modeling with  two mode\-rate-size sunspots, one observed on October 10 2011 in AR 11312 and another observed on March 30 2016 in AR 12525. They both had a penumbral diameter of $\sim45$\arcsec\ (Figure \ref{otherspots}); the size of the associated microwave sources was near the instrumental resolution (see plots of source and beam size at the top of the left column in Figures \ref{modobsOct10} and \ref{modobsMar30}), therefore we could not reliably determine their brightness temperature. The sunspot-associated sources did not show any structure in the RATAN-600 and NoRH observations.

In both sunspots the magnetic field was rather weak, as evidenced from the fact that there was little e-mode and practically no o-mode emission at high frequencies. As a matter of fact, the NoRH o-mode images (R polarization for October 1011 and L for March 2016) show brightness depressions at the location of the spot ({\it c.f.} \citealp{2010ARep...54...69T, 2011LatJP..48...56B, 2015SoPh..290...21R, 2015CosRe..53...10B}). Sunspot-associated o-mode emission (2nd harmonic)  started at at $\sim$11.9\,GHz on October 2011 and $\sim11.5$\,GHz on March 2016, which puts the magnetic field intensity at the base of the TR to $\sim2150$ and $\sim2100$\,G respectively. The potential extrapolations of the HMI longitudinal magnetic field gave maximum intensities of 2230 and 2390\,G respectively at the photosphere and 1700 and 1800\,G at a height of 1000\,km. Thus, either the TR started below 1000\,km or HMI saturated at high $B$ values; we dismissed the first possibility and multiplied the magnetic field by a factor of 1.2 to compensate for saturation effects and keep the base of the TR at reasonable heights. 

\begin{figure}[h]
\begin{center}
\includegraphics[width=0.4\textwidth]{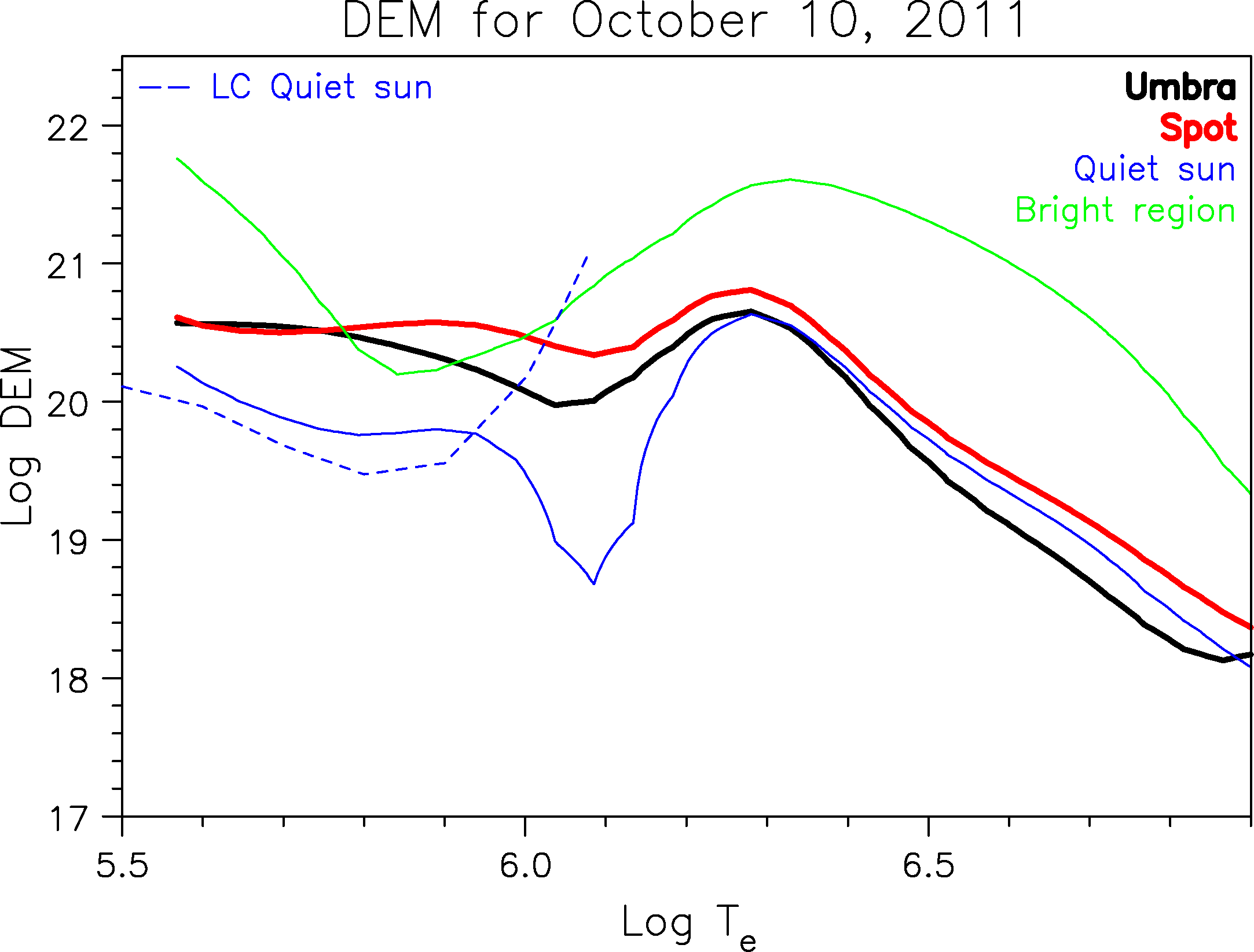}~~~%
\includegraphics[width=0.4\textwidth]{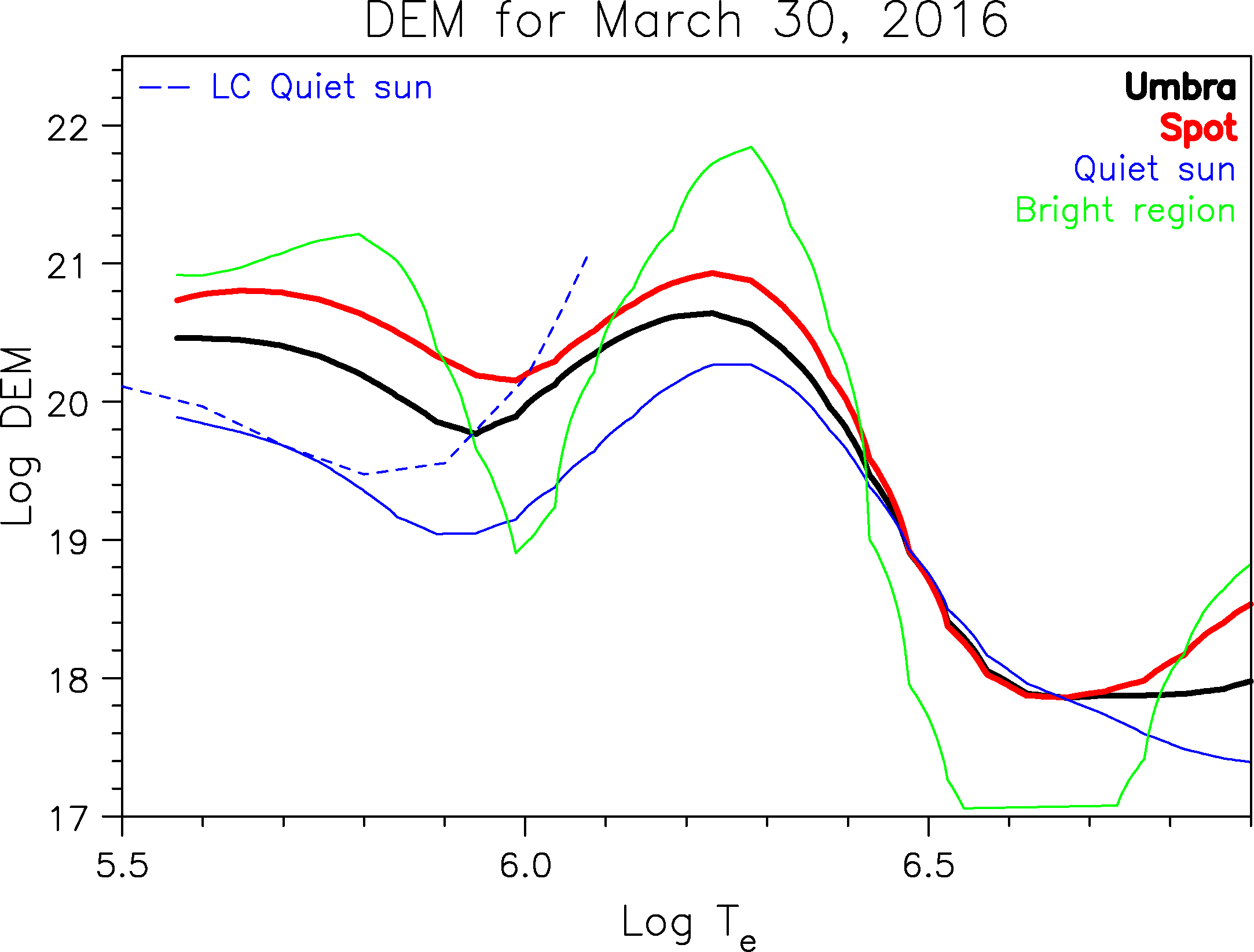}
\end{center}
\caption{Measured DEM in cm$^{-5}$ for October 10, 2011 (left) and March 30, 2016 (right) for $\log T_e<6.9$.}
\label{otherDEM}
\end{figure}

Figure \ref{otherDEM} shows the averaged DEM measured over 4 regions, excluding the high-temperature peak. They are qualitatively similar to our measurements for April 2016 (Figure \ref{EM0}).
Proceeding as described in Section \ref{radiomodels}, we computed best fits for models 1 and 2. For both sunspots the same parameters gave good fits for both R and L polarizations, as well as for models 1a and 1b; their values are given in Table \ref{modpar2}.

\begin{figure}[h]
\begin{center}
\includegraphics[height=7.7cm]{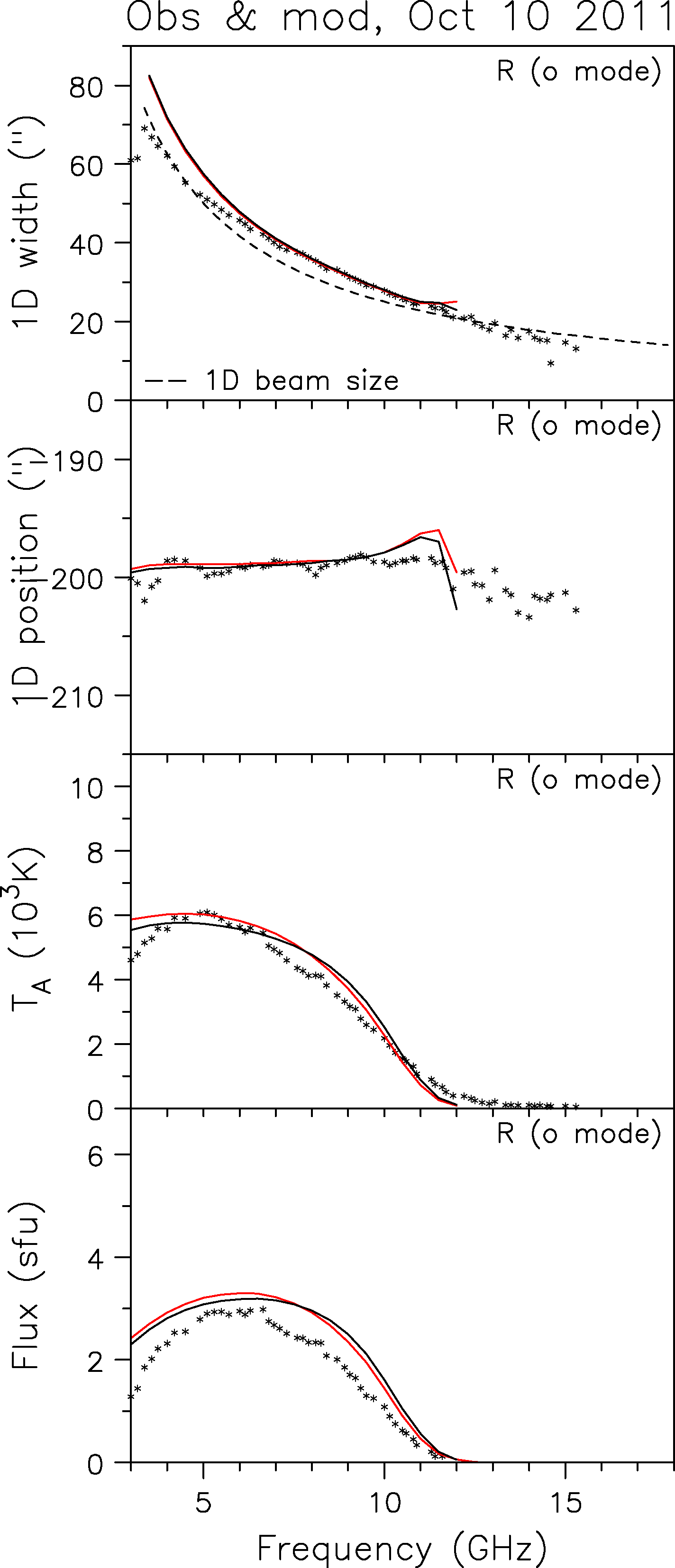}~\includegraphics[height=7.7cm]{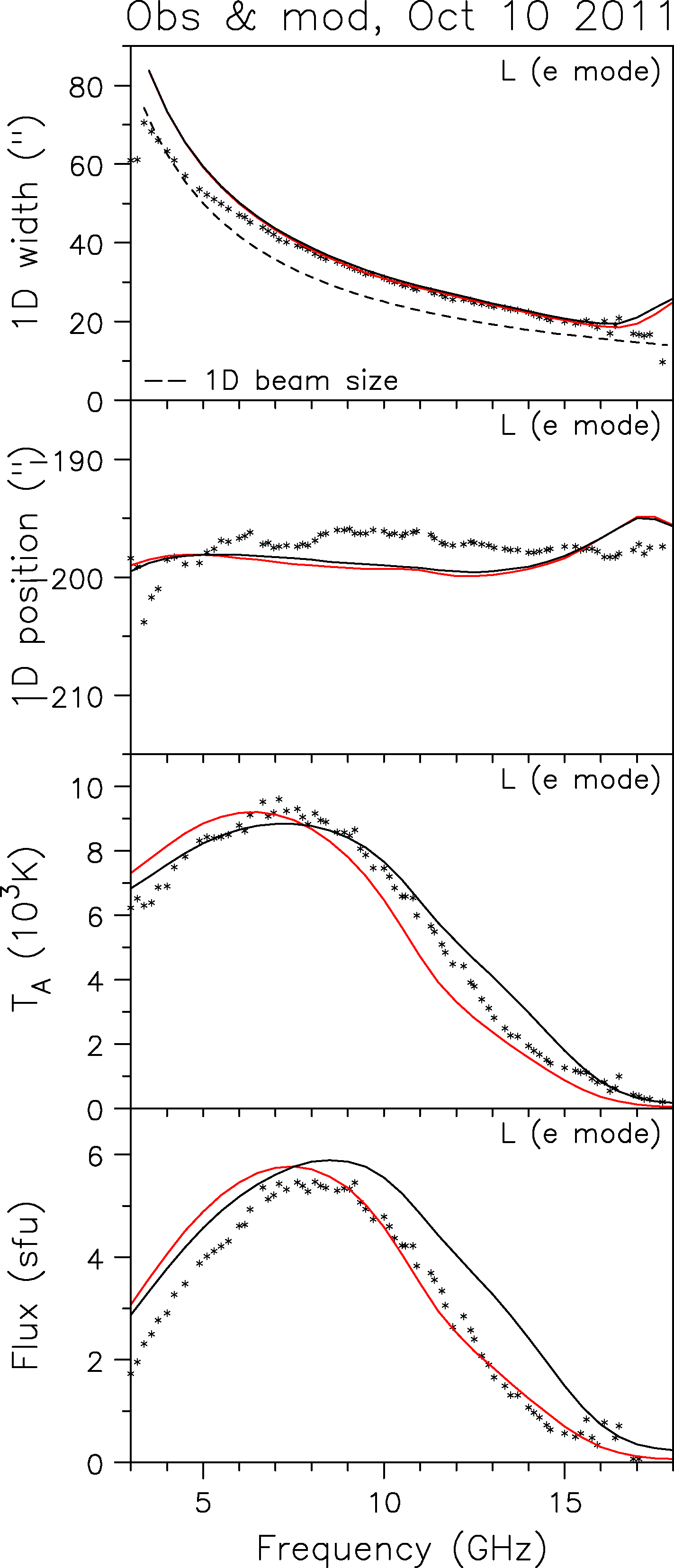}~\includegraphics[height=7.5cm]{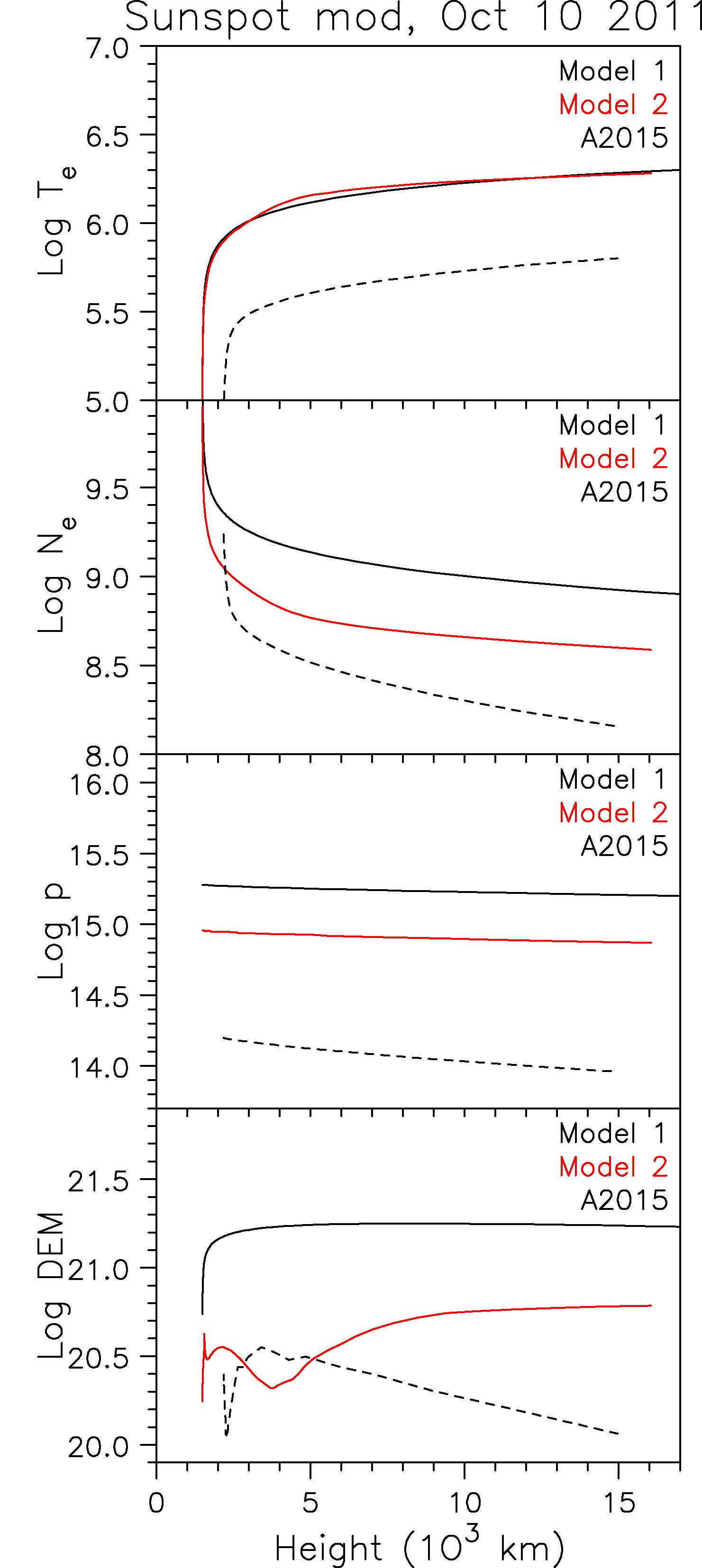}
\end{center}
\caption{Comparison of observed (asterisks) and model (full lines) parameters for R and L polarizations (left and center) for the spot of October 10, 2011. Model 1a is shown in black, model 1b in green and model 2 in red. The dashed line in the width plots gives the instrumental beam width. The plots at right show the physical parameters as a function of height for the three models, as well as for the sunspot model of \citealp{2015ApJ...811...87A} (dashed lines).}
\label{modobsOct10}
\end{figure}
\begin{figure}[!h]
\begin{center}
\includegraphics[height=7.7cm]{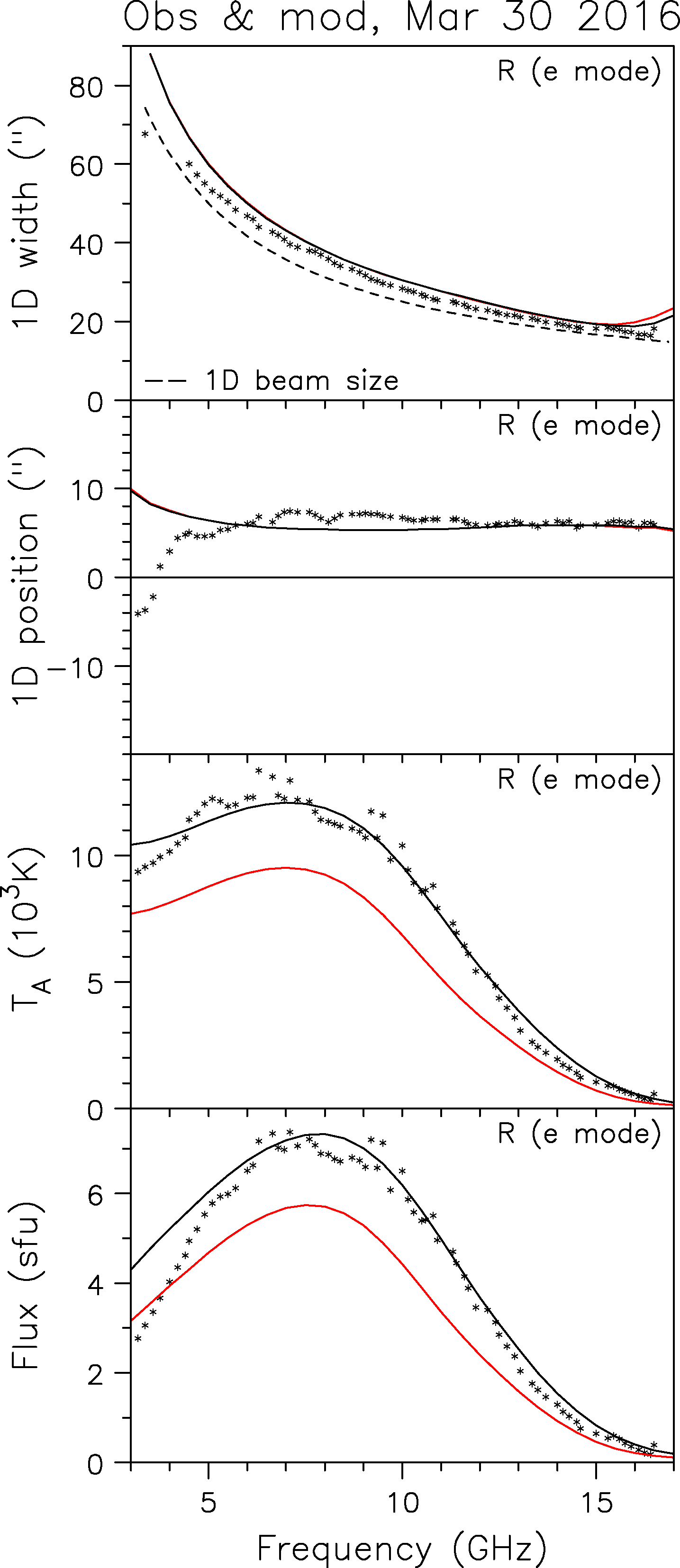}~\includegraphics[height=7.7cm]{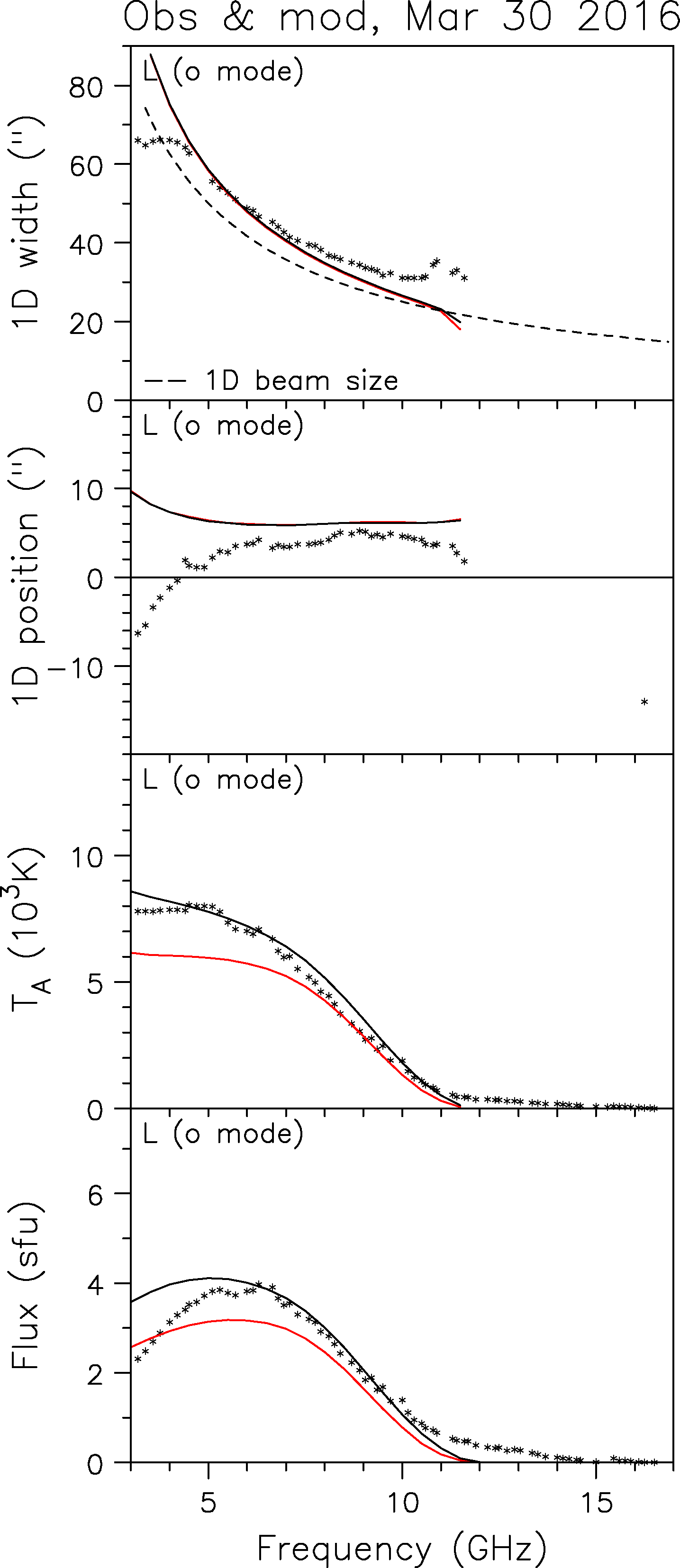}~\includegraphics[height=7.5cm]{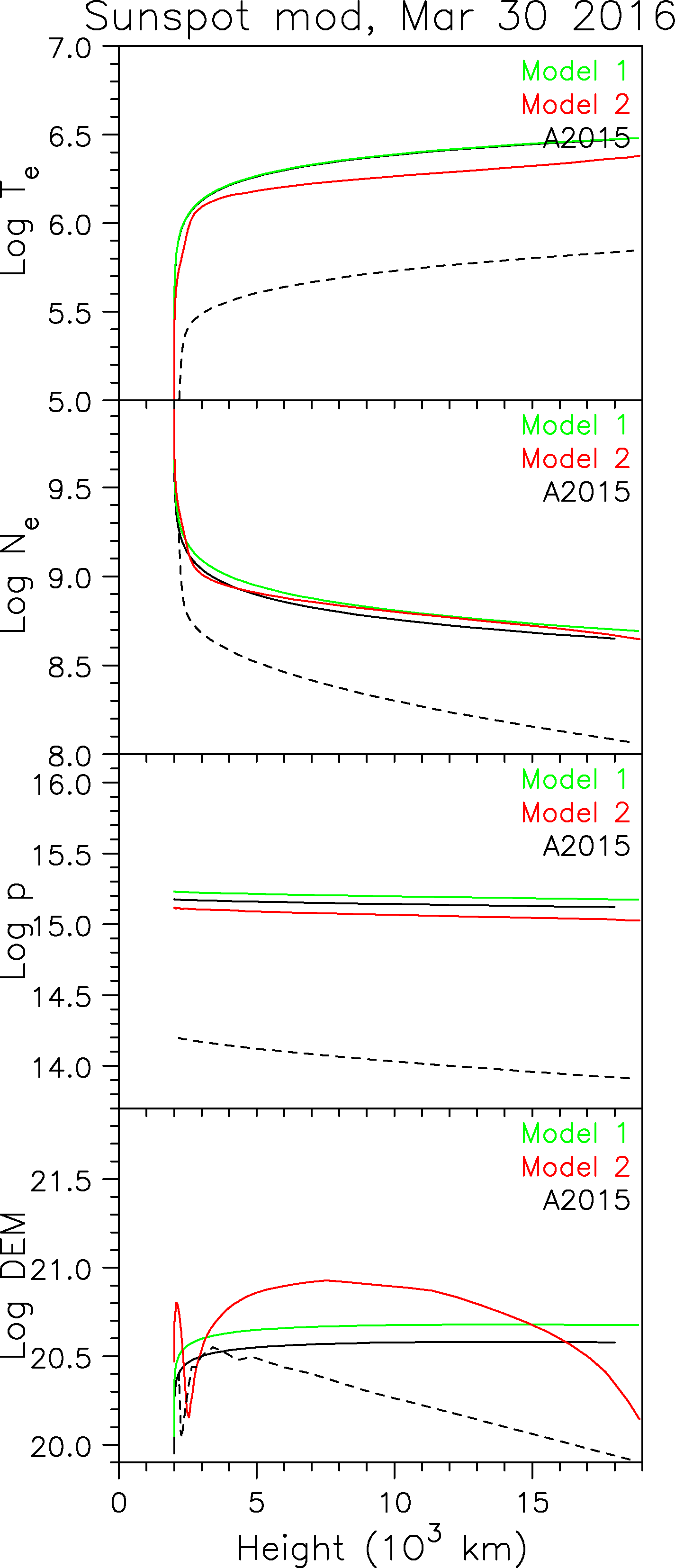}
\end{center}
\caption{Same as Figure \ref{modobsOct10} for the spot of March 30, 2016 and for models 1 and 2.}
\label{modobsMar30}
\end{figure}

\begin{table}[h]
\caption{ Model parameters for the two moderate-size sunspots}
\label{modpar2}
\begin{tabular}{l|rr|rr}
\hline 
&\multicolumn{2}{c}{October 2011}&\multicolumn{2}{c}{March 2016}\\
Parameter                            & Mod 1& Mod 2& Mod 1& Mod 2\\
\hline 
$H_0$~(km)                           & 1500 & 1500 & 2000 & 2000 \\
$p_0~(10^{15}$\,K\, cm$^{-3})$       & 1.9  & 0.9  & 1.7  & 1.3  \\
$F_c~(10^6$\,erg\,cm$^{-2}$s$^{-1}$) & 2.3  & --   & 9.0  & --   \\
\hline 
\end{tabular}
\end{table}    

\begin{figure}[ht]
\begin{center}
\includegraphics[width=0.47\textwidth]{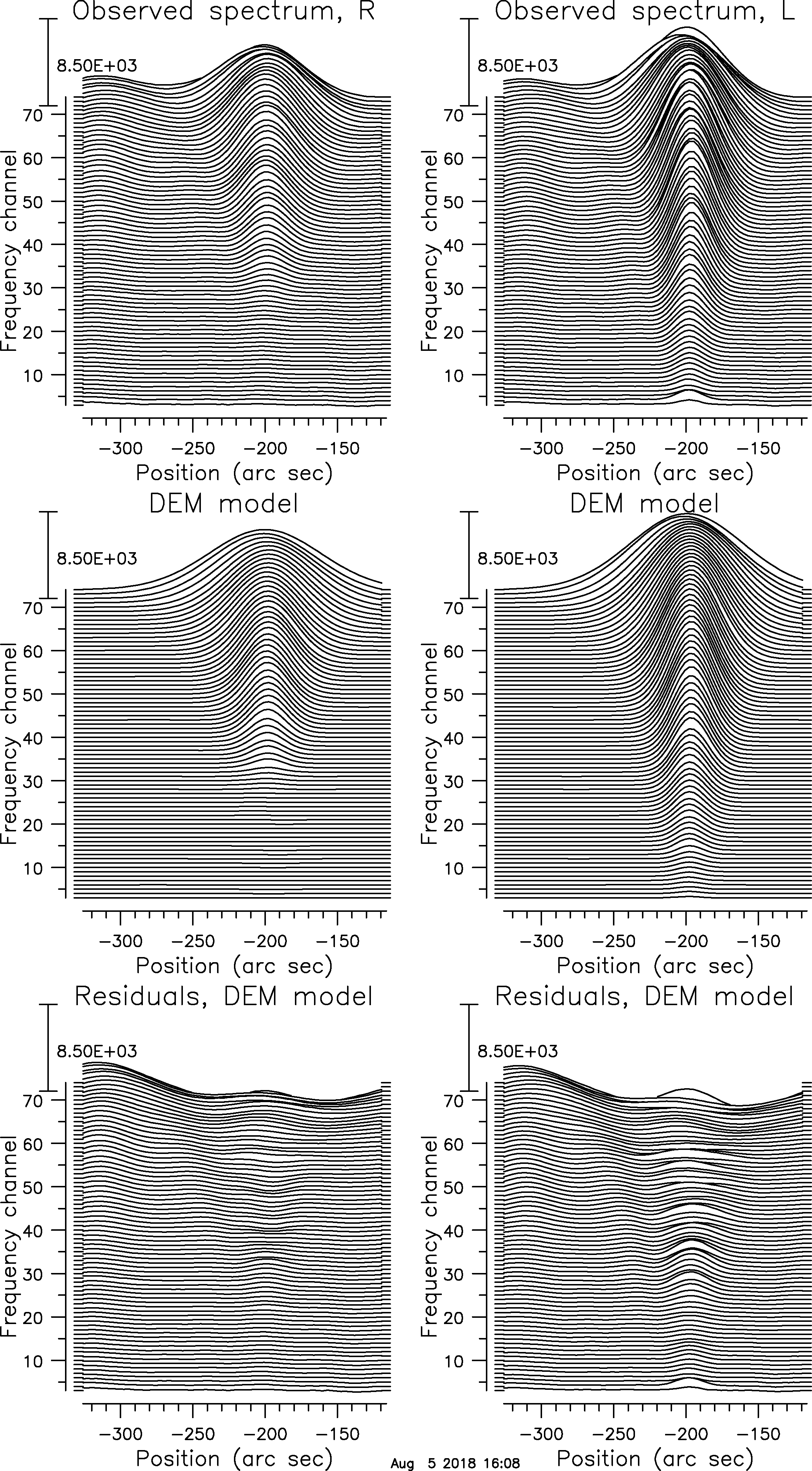}%
\hspace{0.7cm}%
\includegraphics[width=0.47\textwidth]{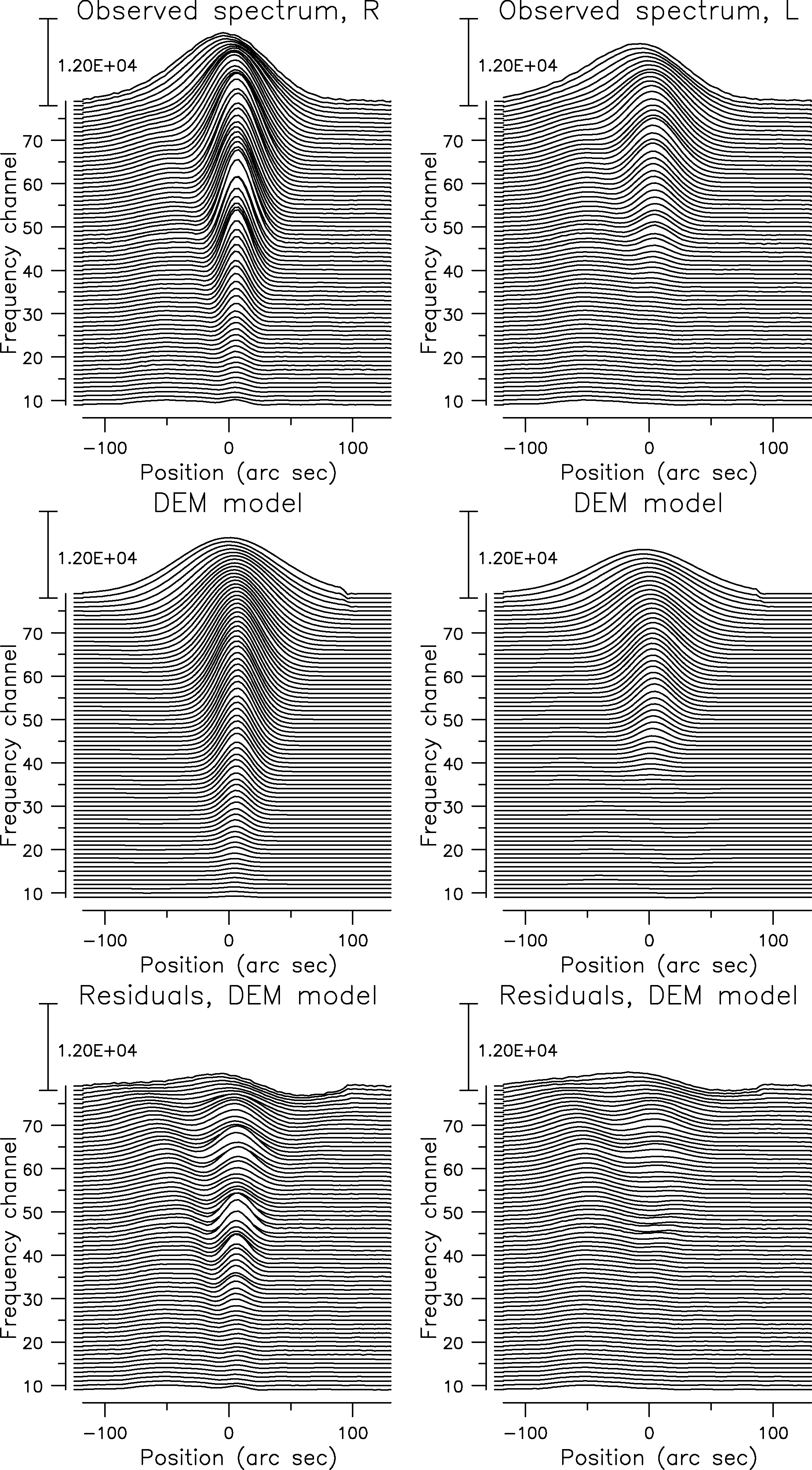}
\end{center}
\caption{Observed RATAN-600 1D spectra and the corresponding DEM models and residuals for the sunspots in active regions 11312 {(left)} and 12526 (right). The emission East of the sunspot in AR 11312 is plage-associated}.
\label{othermods}
\end{figure}

We present the results for October 2011 in Figure \ref{modobsOct10}. In this case the value of the pressure parameter for Model 2 is below the maximum possible value. Both the constant $F_c$ and the DEM models reproduce fairly well the observations, with the DEM model antenna temperature differing from the observed by 3\% (RCP) and 9\% (LCP) on the average. Their temperature structure is very similar, while the density (and pressure) predicted by model 2 is lower than that of model 1 (right column of plots in Figure \ref{modobsOct10}); this density difference does not affect appreciably either the peak antenna temperature or the flux, apparently because both the 2nd and the 3rd harmonic are already optically thick.  The DEM model and the residuals, together with the observed 1D spectra are shown in Figure \ref{othermods} (right).

The results for March 2016 are shown in Figure \ref{modobsMar30}. Model 1 (black lines) reproduces fairly well all observed parameters.  For model 2 (red lines) we had to use the highest possible value for $p_0$; on the average, model 2 $T_A$ is 27\% below the observed for RCP and 21\% for LCP. The difference is because model 2 predicts lower temperature than model 1 at large heights, where the low frequency emission originates (top plot in the right column of Figure \ref{modobsMar30}). The two models are very close in $N_e$, while both $T_e$ and $N_e$ are above the  \cite{2015ApJ...811...87A} sunspot model (right column of plots in Figure \ref{modobsMar30}). The DEM model and the residuals, together with the observed 1D spectra are shown in Figure \ref{othermods} (left).

We conclude from this analysis that the $T_e$-$N_e$ structure derived from the inversion of the DEM is broadly consistent with the microwave observations of these spots. 

\subsection{The large sunspot of April 14, 2016}\label{Apr16}
In this section we focus on NOAA active region 12529, which traversed the solar disk from April 7 to April 20, 2016 and contained a very large leading spot that crossed the central meridian on April 13, $\sim$22 UT. The penumbral diameter of the spot was 80\arcsec\ and the umbral diameter 50\arcsec. This particular sunspot attracted our attention because its microwave emission showed two resolved components, visible both in the RATAN-600 and NoRH data. We selected for further study the observations on April 14. 

The availability of EUV data gave us a good opportunity to check whether this structure is due to temperature/density variations across the spot,  or is due to the variation of the angle between the magnetic field and the line of sight.  Thus, after the presentation of the observations, we discuss the DEM variation across the sunspot and  then proceed with the modeling of the RATAN-600 and the NoRH emission.

\begin{figure}[!h]
\begin{center}
\includegraphics[width=0.55\textwidth]{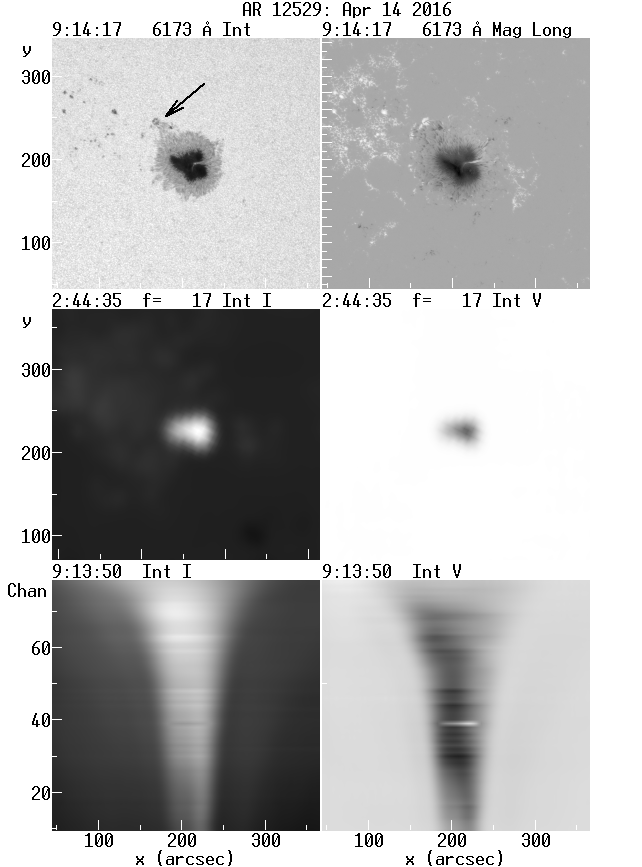}
\end{center}
\caption{Observations of the leading spot in AR 12529. Top: Continuum intensity and longitudinal magnetic field from HMI. Middle: NoRH images in Stokes I and V at 17\,GHz (1.7\,cm). The peak antenna temperatures are $6.68\times10^5$\,K for I and $-5.44\times10^5$ for V. Bottom: RATAN-600 one-dimensional spectral images in $I$ and $V$; the spectral range is from 16.5 GHz (1.82\,cm) at the bottom to 3.37\,GHz (8.90\,cm) at the top. The arrow in the upper left panel points to a moving pore, discussed in Section \ref{Model_RATAN}. Images are oriented with the celestial North up.}
\label{overview}
\end{figure}

\subsubsection{Overview of the observations}
On April 14, 2016 the RATAN-600 data were obtained around 09:14 UT in 69 spectral channels, from 16.5 to 3.37 GHz.
Figure \ref{overview} shows images and spectra of the spot. We note that the NoRH images show two components in both $I$ and $V$. The 17\,GHz source is highly polarized, $\sim80$\% which, for gyro-resonance emission, shows that radiation comes from the 3rd harmonic of the gyrofrequency, being optically thick in the extraordinary mode (e-mode) and thin in the ordinary (o-mode). The brightness temperature of the e-mode emission (left circularly polarized here, corresponding to the negative magnetic field of the sunspot), $1.21\times10^6$\,K, is a lower limit to the electron temperature at the height where the intensity of the magnetic field is $B=2040$\,G; this temperature corresponds to the upper chromosphere-corona transition region (TR). We add that the absence of emission at 34\,GHz shows that $B<4080$\,G at the base of the TR.
 
\begin{figure}[ht]
\begin{center}
\includegraphics[height=7cm]{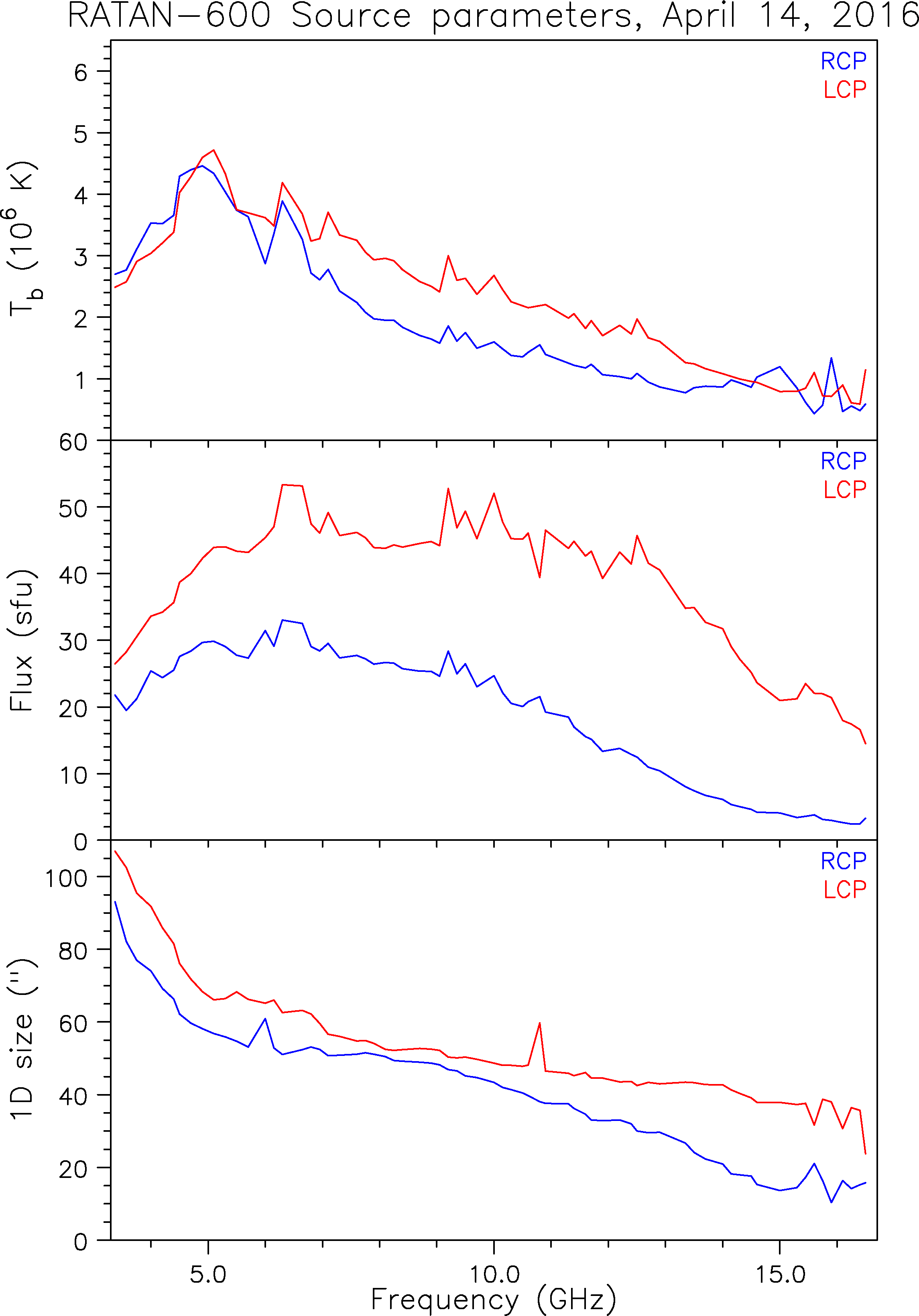}~~~\includegraphics[height=7cm]{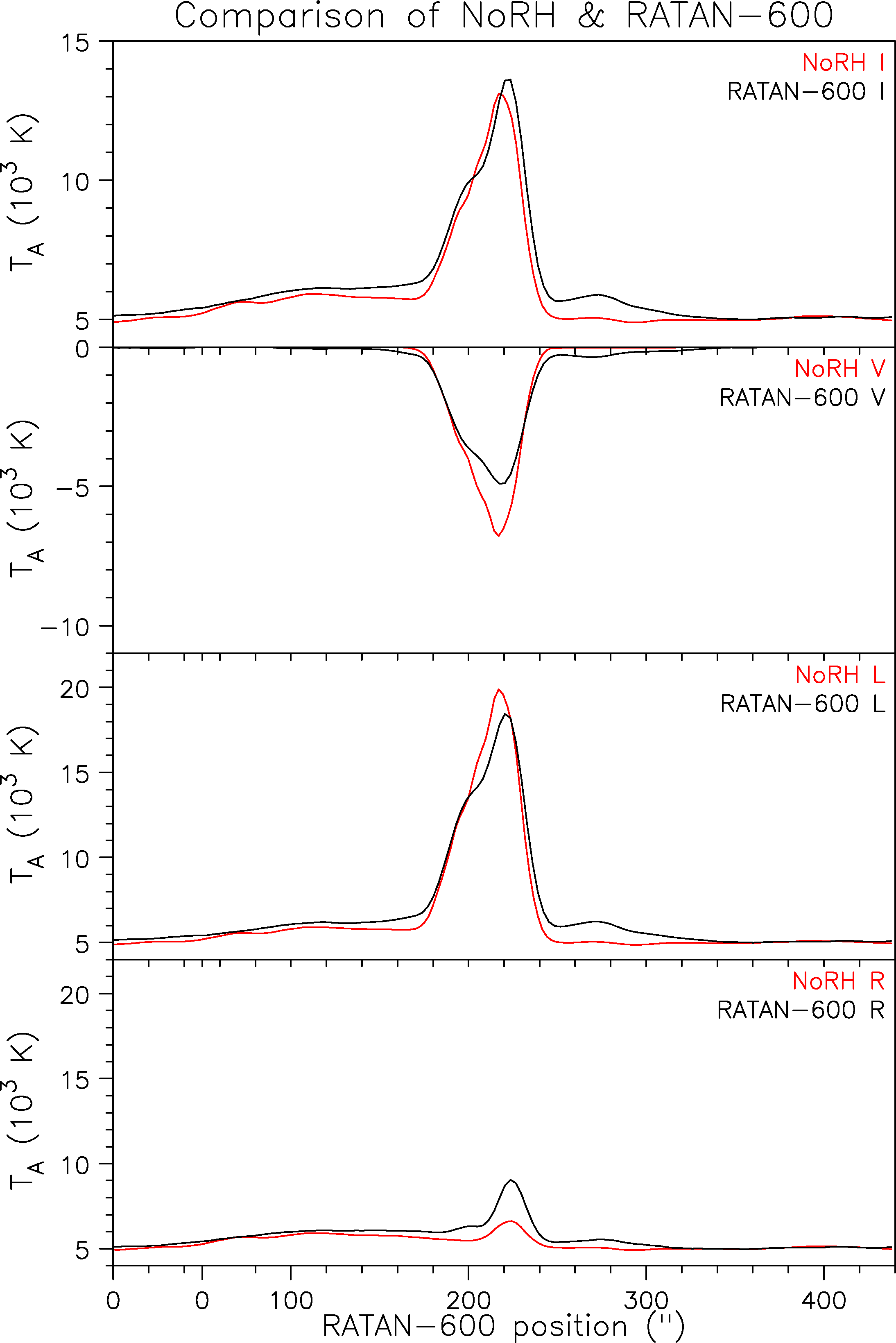}
\end{center}
\caption{Left panel: Size (corrected for instrumental resolution), flux and brightness temperature of the sunspot-associated source as a function of frequency, from RATAN-600 data. Right panel: comparison of the RATAN-600 1D image at 16.5\,GHz, 09:14 UT, with the corresponding NoRH data at 17\,GHz, 02:45 UT. $T_A$ is the RATAN-600 antenna temperature in K; { the conversion factor from $T_A$ to sfu/\arcsec is $3.63\times10^{-5}$}. The NoRH plots have been shifted by 50\arcsec\ to compensate for solar rotation. The scales for I and V are not the same as those for R and L.}
\label{parm_comp}
\end{figure}

The RATAN-600 1D images also show two components in the highest frequency channels, where the resolution is higher. The brightness temperature can be estimated from the 1D scans under the assumption of a Gaussian-shaped circular source. This requires the observed width is not too close to the beam width, in order for the true width to be computed reliably, and this condition is satisfied for this particular source. Due to the double-peaked structure of the source this approximation is not accurate here for the high frequencies, but it should be satisfactory for intermediate and low frequencies, where we expect the source to be more flat. 

The results are shown in the left panel of Figure  \ref{parm_comp}. We note that the brightness temperature increases at longer wavelengths, reaching a maximum of $4.7\times10^6$\,K in the e-mode at $\lambda=5.9$\,cm (5.1\,GHz). In terms of the gr mechanism, this implies that (a) the electron temperature as at least that high in the corona above the spot and (b) the magnetic field at that height is 620\,G (corresponding to the 3rd harmonic of the gyrofrequency), or 465\,G (4th harmonic). Figure 2 also shows that the size of the source increases with wavelength. As the values have been corrected for the effects of the instrumental beam width, the increase of the size reflects the increase of the diameter of the harmonic layers as they move higher with increased wavelength.

In the right panel of Figure \ref{parm_comp} we compare the NoRH data at 17\,GHz with the RATAN-600 data at the highest available frequency, 16.5\,GHz. For this purpose we integrated the NoRH image in the celestial NS direction, taking into account the effect of the RATAN-600 NS beam and expressed the results in terms of RATAN-600 antenna temperature, $T_A$.  We note that the two data sets compare fairly well as far as the quiet sun background and the total intensity are concerned, with the NoRH giving a somewhat smaller value of circular polarization. We conclude that the two data sets are to a satisfactory degree consistent with each other.

\subsubsection{Horizontal structure of the spot}
\begin{figure}[h]
\begin{center}
\includegraphics[width=.45\textwidth]{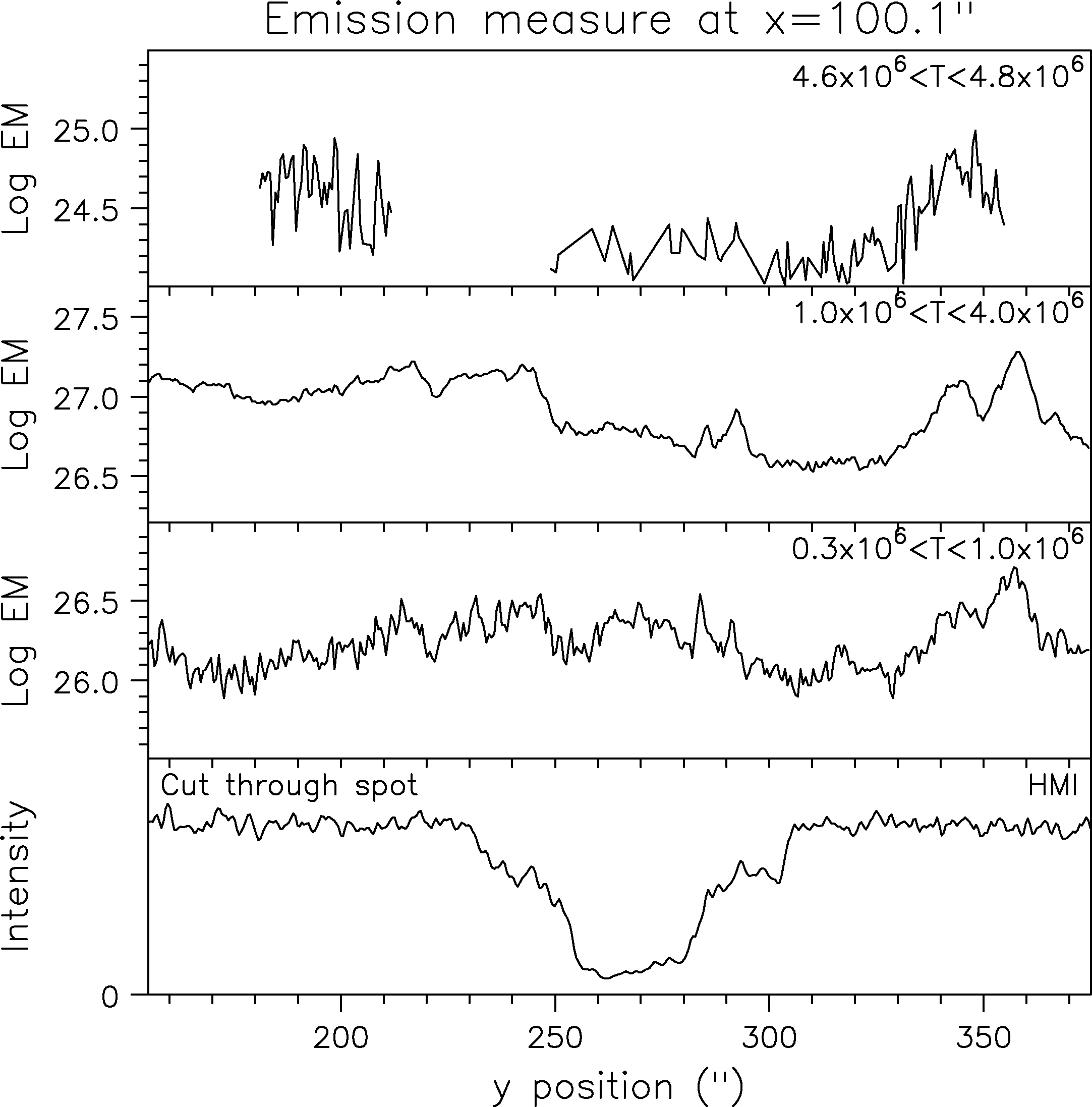}
\end{center}
\caption{The emission measure in cm$^{-5}$ along the cut of Figure \ref{dem} for three temperature ranges. The bottom plot shows the intensity across the spot.}
\label{EM}
\end{figure}

The DEM cut at the right of Figure \ref{dem} shows that there is no appreciable temperature difference between the spot and the rest of the active region. This is not the case with the density, as evidenced from the DEM images and the cut in Figure \ref{dem} and better shown in Figure \ref{EM}, where the emission measure integrated over three temperature ranges is plotted across the spot. In both the 0.3-1.0$\times10^6$\,K and the  1.0-4.0$\times10^6$\,K ranges, there is a drop of EM between the southern penumbra and the umbra of 0.25 and 0.35 in log EM, also visible in Figure \ref{dem}; this corresponds roughly to a density decrease by factors of 1.3 and 1.5 respectively, which is consistent with our estimate in Section \ref{DEM_AIA}. We also note here, that EM/DEM differences may not only reflect density differences, but formation length differences as well. However, at the N side of the spot and further in the quiet sun, we can see no marked EM increase.

We conclude that the temperature/density variation across the spot is not sufficient to explain the double-peaked structure in the microwave range. We also note that there is very little plasma around 4.7$\times16^6$\,K (top plot in Figure \ref{EM}, {\it c.f.} also the cut in Figure \ref{dem}), which is required in order to produce the maximum observed $T_b$ though thermal emission.

\subsubsection{Modeling of the RATAN-600 observations}\label{Model_RATAN}
For the April 2016 spot the HMI magnetograms were saturated, due to the high $B$ value and hence unusable; moreover, no magnetograms were available from Hinode/SOT. However, a scan of the entire active region with the Hinode/SOT spectropolarimeter was available, obtained from 02:21 to 03:24 on the day of our observations. The longitudinal magnetic field computed from this scan showed intensities as high as $-3750$\,G. Another interesting aspect is that the partial light bridge in the NW part of the spot (Figure \ref{overview}) was of opposite polarity, with $B\sim2400$\,G above the penumbra and $\sim1000$\,G well within the umbra. We note that this region was the site of continuous activity, very well depicted in two IRIS movies, from 04:27 to 05:27 and from 05:40 to 06:40 UT, available at the IRIS site.

\begin{figure}[h]
\begin{center}
\includegraphics[width=\textwidth]{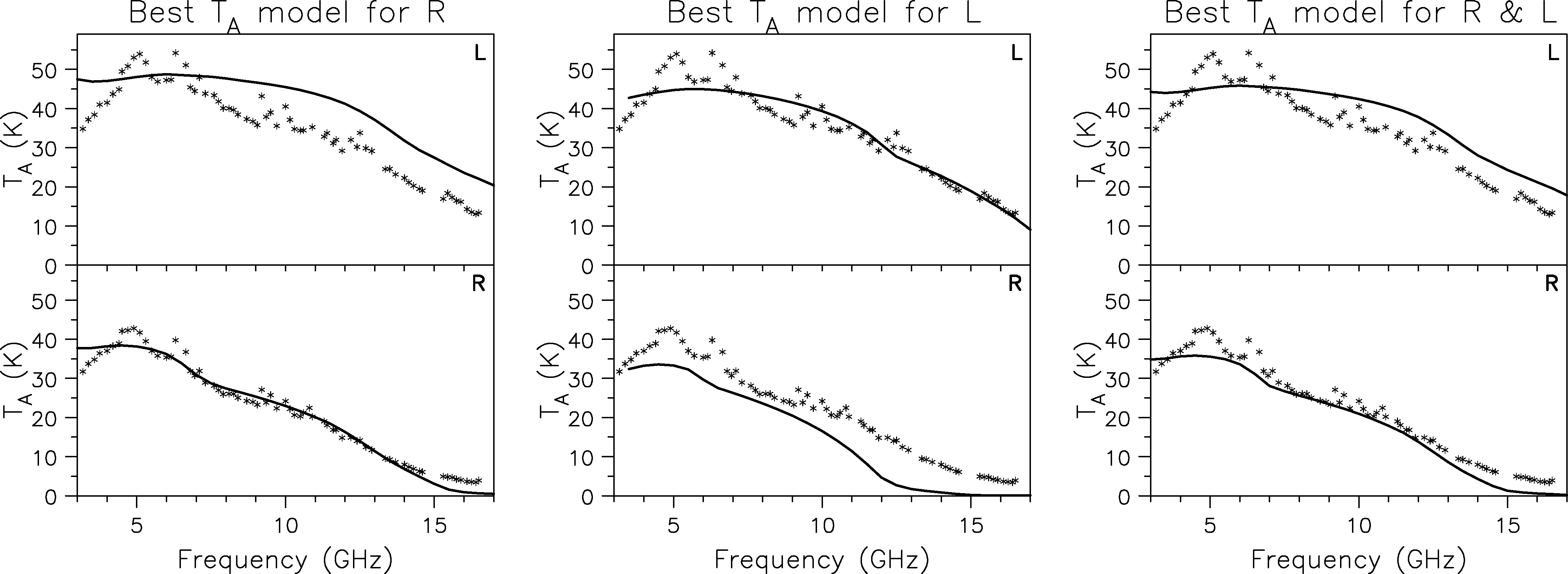}
\end{center}
\caption{Observed (asterisks) and model (full lines) peak RATAN-600 antenna temperature as a function of frequency for the best R model, the best L model and the best combined RL model.}
\label{modcomp}
\end{figure}

The observed peak $T_A$ spectra together with the best results from Model 1, method a,  are shown in Figure \ref{modcomp}, for both L (e-mode) and R (o-mode) polarization. The models were computed with the best fit parameters for R (left panel), L (middle) and combined RL (right). Although the agreement between the observed and model $T_A$ is not perfect, the two are reasonable close to each other for the entire frequency range; this is also true for the combined RL model, for which the average deviation ($\sqrt{\chi_{RL}^2}$) is 6.6\%. It is noteworthy that the largest deviation is around 5\,MHz, where the maximum $T_b$ was measured (section \ref{radobser}). Actually, the maximum $T_b$ predicted by the model is 3.8$\times10^6$\.K, compared to the observed value of 4.7$\times10^6$\.K.

\begin{figure}[h]
\begin{center}
\includegraphics[height=4.5cm]{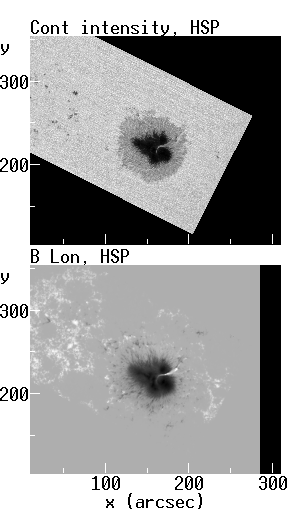}\,\includegraphics[height=4.5cm]{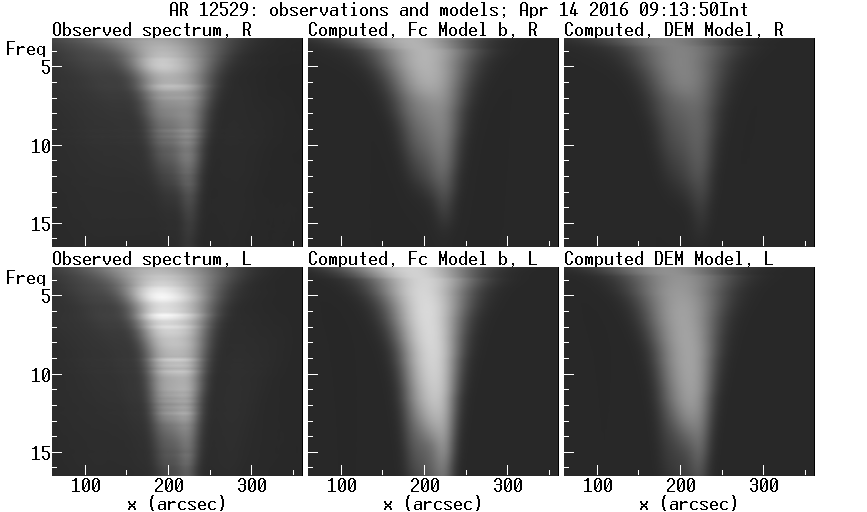}
\end{center}
\caption{Observed RATAN-600 1D spectra and computed spectra from the $F_c$ and the DEM models. The intensity scale is linear and the same for all displays, to facilitate the comparison. The continuum intensity and $B_\ell$ images are given in the left column for reference.}
\label{obsmod}
\end{figure}

\begin{figure}[h]
\begin{center}
\includegraphics[width=0.7\textwidth]{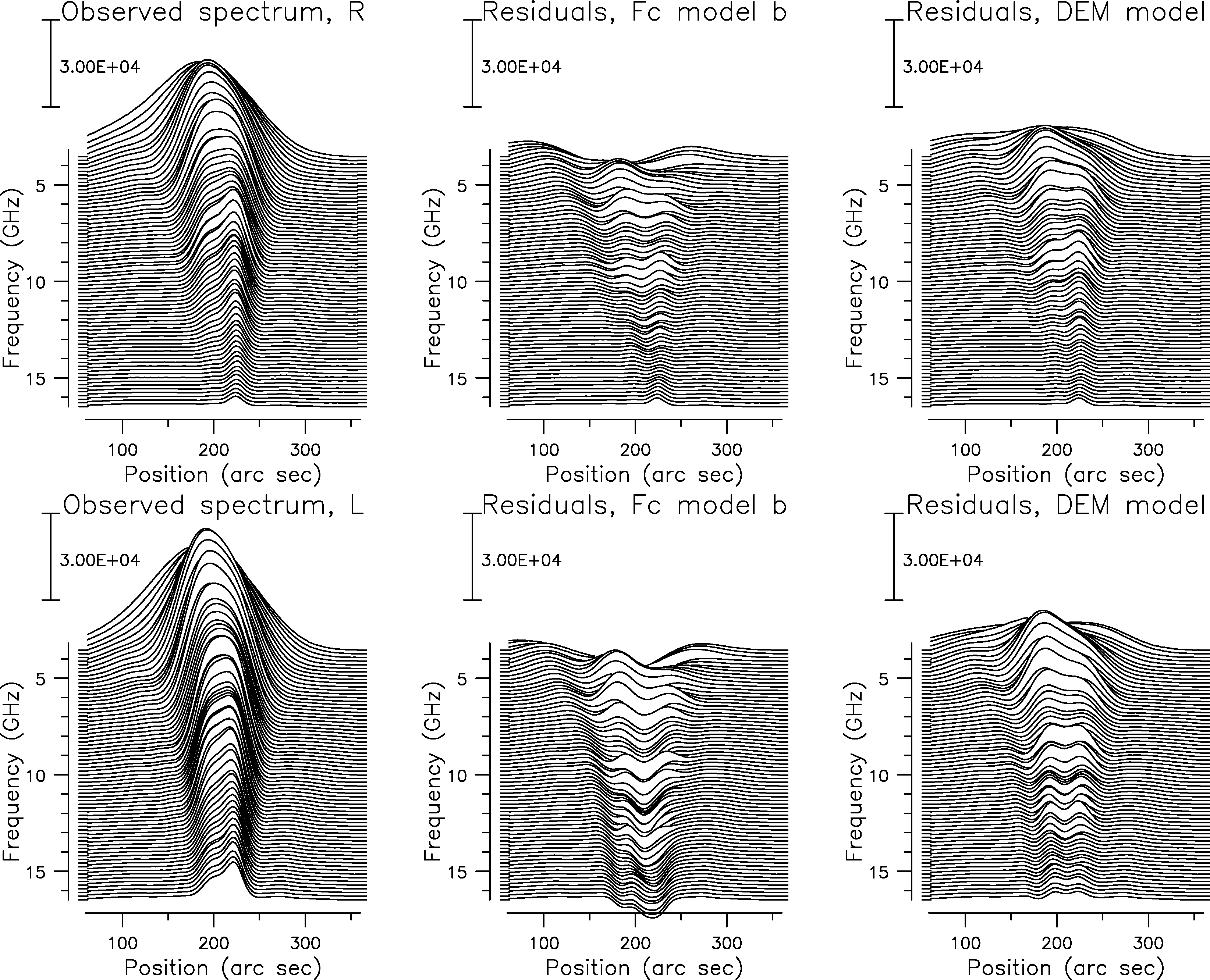}
\end{center}
\caption{{ Tracings of the observed RATAN-600 1D spectra and the residuals for the $F_c$ and DEM models. The intensity scale (vertical bar) is the same for all plots. The short horizontal segments at either side of the tracings mark the zero level.}}
\label{resid}
\end{figure}

The computed antenna temperature for Model 1b (best combined RL), together with the observed RATAN-600 R and L spectra and the Hinode SP observations are presented in Figure \ref{obsmod}, while the residuals are plotted in Figure \ref{resid}. The
values of the model parameters for best R, best L and best combined RL models are given in Table \ref{modpar1}. In addition to the strong similarity of model and observed 1D spectra, we note that at high frequencies the model reproduces fairly well the double-peaked structure seen in the RATAN-600 scans. This implies that its origin is due to the small angle of the magnetic field with respect to the line of sight above the penumbra and hence the variation of $N_e$ or $T_e$ across the sunspot have little effect, if any. 

\begin{table}[h]
\caption{Model parameters for AR 12529 (April 14, 2016)}
\label{modpar1}
\begin{tabular}{l|rrr|rrr|r}
\hline 
&\multicolumn{3}{c}{$F_c$ model, a}&\multicolumn{3}{c}{$F_c$ model, b}&DEM\\
Parameter                            & R    & L     & RL   & R    & L    & RL   & RL   \\
\hline 
$H_0$~(km)                           & 1600 &  3500 & 1900 & 1200 & 2800 & 1500 & 1000 \\
$p_0~(10^{15}$\,K\, cm$^{-3})$       & 3.8  &   7.4 &  4.3 &  3.2 &  7.5 &  2.8 & 1.1  \\
$F_c~(10^6$\,erg\,cm$^{-2}$s$^{-1}$) & 11.7 &  12.3 & 10.3 & 10.6 & 13.0 &  9.4 &  --  \\
\hline 
\end{tabular}
\end{table}    

Both $F_c$ models give small residuals, although the corresponding parameters are not identical. We note here that the three model parameters are not completely independent: once the 2nd and 3rd harmonic layers are both in the TR, an increase of $H_0$ can be compensated by a steepening of the temperature gradient. The DEM model gave greater residuals, and antenna temperatures $\sim30$\% below the observed, considerably larger than for the moderate-size spots discussed in Section \ref{small_spots}.

\begin{figure}[h]
\begin{center}
\includegraphics[height=9.0cm]{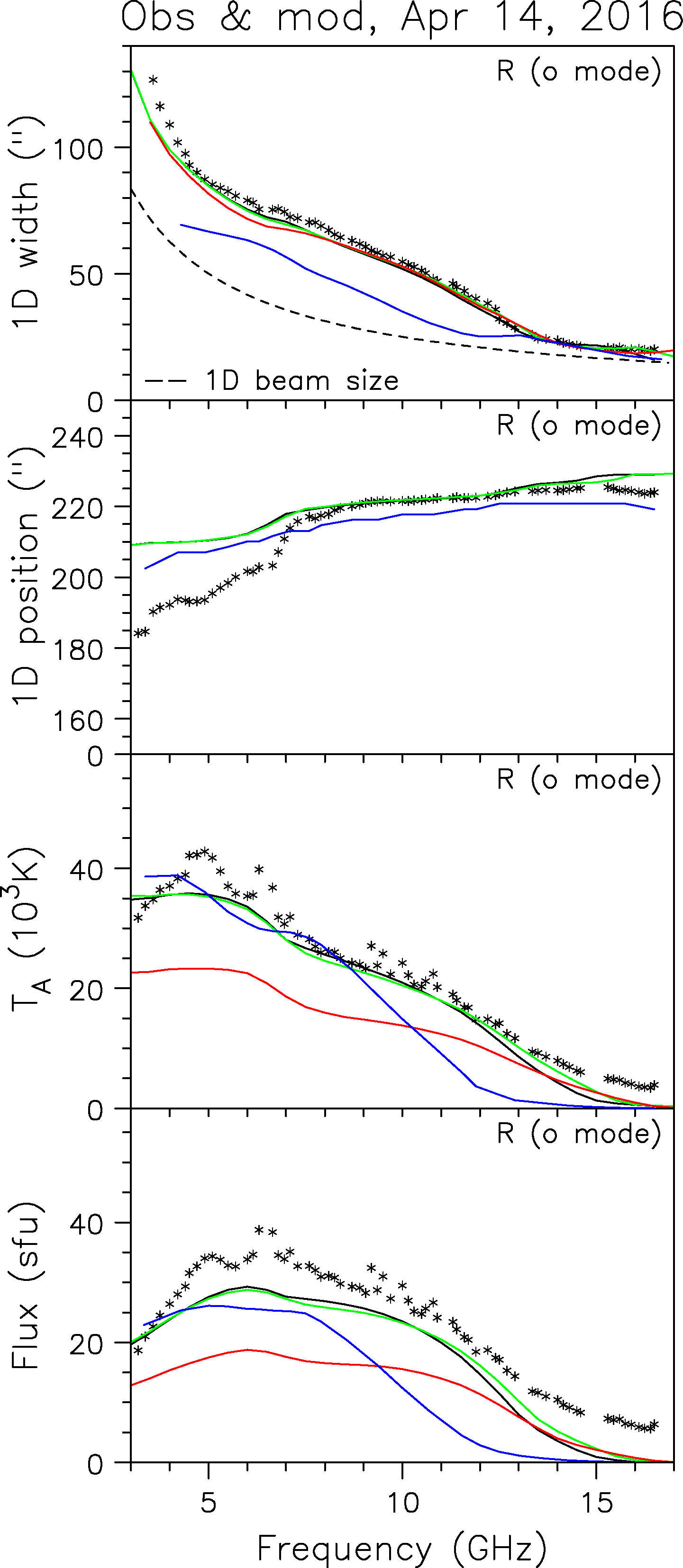}~\includegraphics[height=9.0cm]{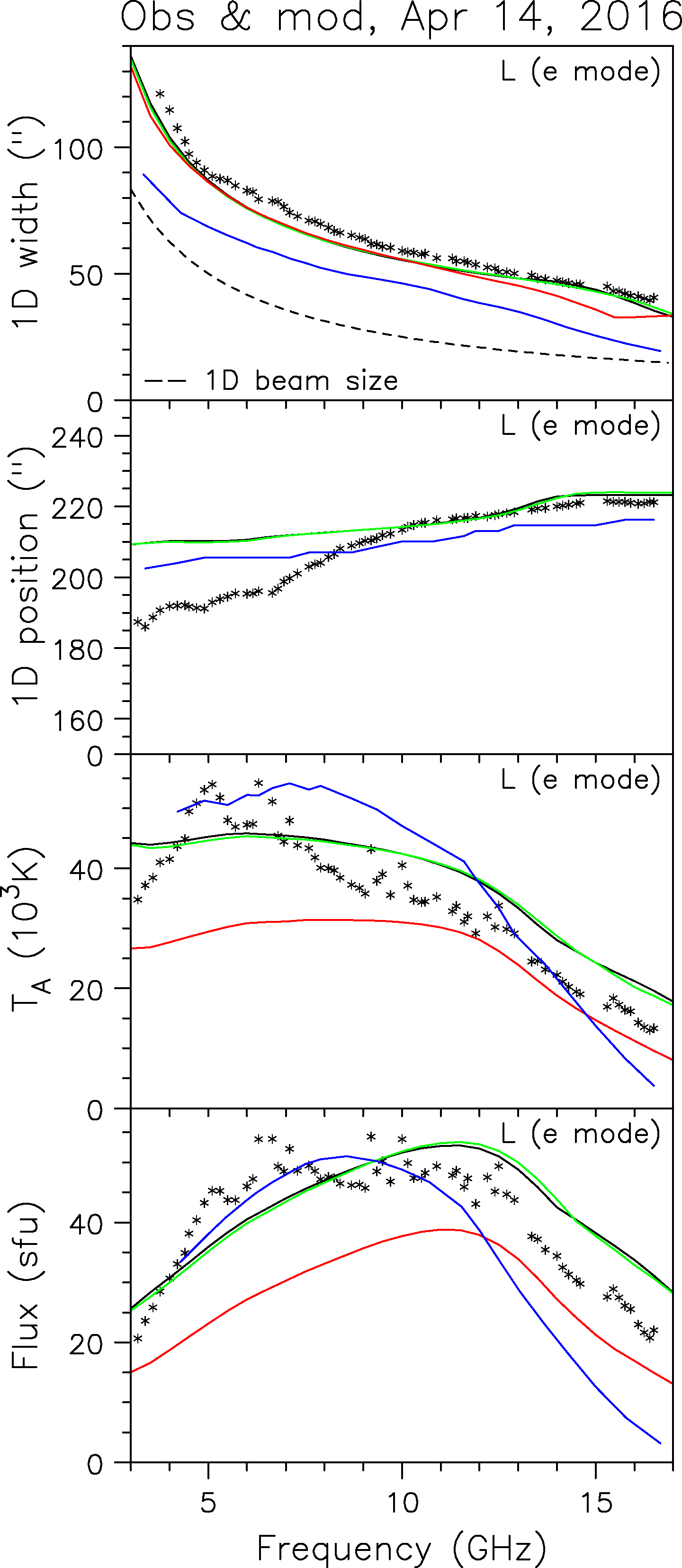}~\includegraphics[height=9.0cm]{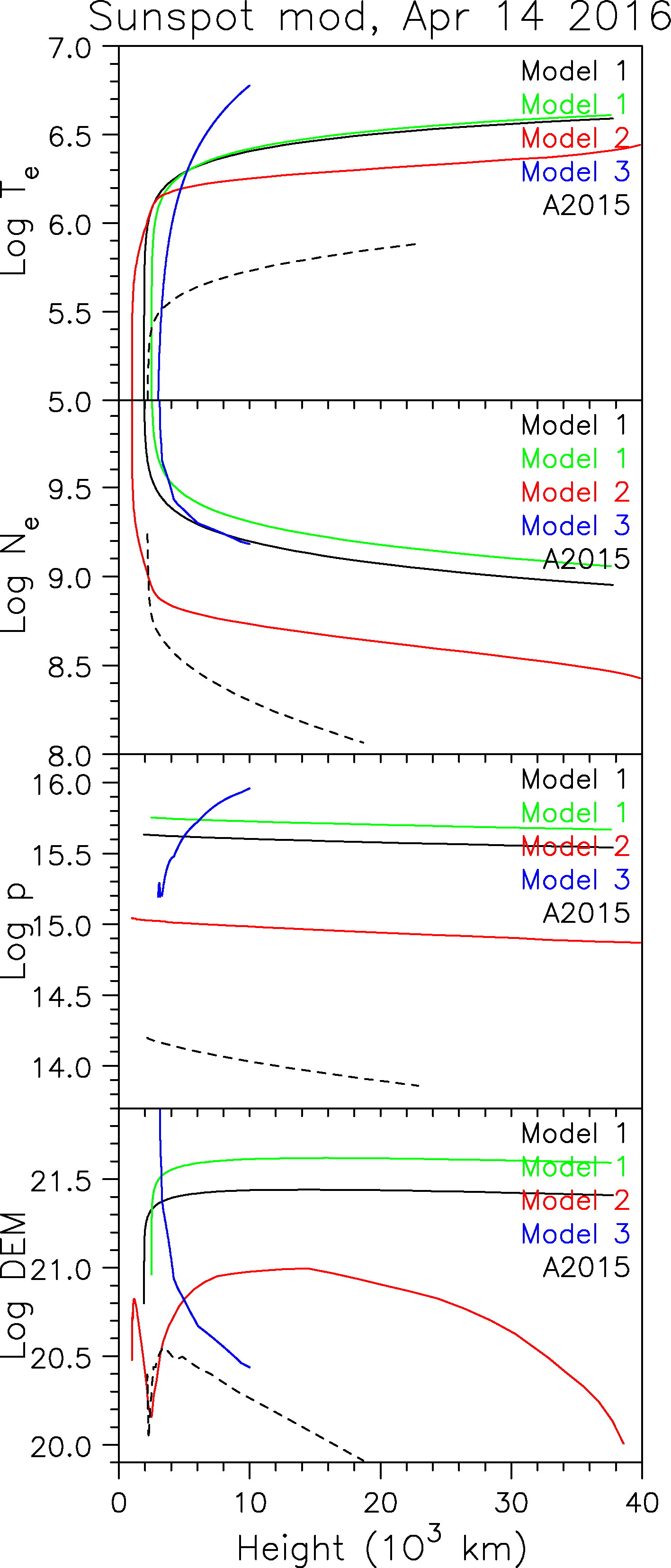}
\end{center}
\caption{Comparison of observed (asterisks) and model (full lines) parameters for R and L polarizations (left and center). Model 1a is shown in black, model 1b in green, model 2 in red and model 3 in blue. The dashed line in the width plots gives the instrumental beam width. The plots at right show the physical parameters as a function of height for the three models, as well as for the sunspot model of \citealp{2015ApJ...811...87A} (dashed lines).}
\label{modobsApr16}
\end{figure}

Figure \ref{modobsApr16}, left and center panels, compares all the bulk parameters (flux, peak antenna temperature, size and position) of the source deduced from the models (full lines) with the corresponding observed values (asterisks). The position was corrected for the solar rotation between the time of the magnetogram and the observations. In addition to the constant $F_c$ and DEM models, we used here a third model (Model 3), based on the model of \cite{2008A&A...488.1079S} for active regions, which has been used in the past for modeling the sunspot microwave emission (e.g. \citealp{2015Ge&Ae..55.1124K, 2018SoPh..293...13S}); the density  and the height of TR  were the same, but the coronal temperature was taken equal to $6\times10^6$\,K.

The comparison of the observed and model flux is less satisfactory than that of the antenna temperature, but the observed and model sizes are in very good agreement; we also note that the observed width is well above the beam width, plotted as a dashed line in the same panel. As for the position, we have good agreement at high frequencies and significant deviations at low frequencies. We note that the gradual eastward shift at low frequencies is a geometric effect, due to the fact that at those frequencies the radiation comes from higher layers. The much higher shift of the observed position at low frequencies may indicate that the geometry of the field is very different from the extrapolated. Alternatively, it is possible that another source, besides the sunspot, contributes to the emission at these frequencies.

As evidenced from the figure, models 1a (black lines) and 1b (green lines) give very similar spectra. The DEM model 2 (red lines)
predicts lower than the observed values for both the peak $T_A$ and the flux, as already noted in the discussion of the residuals (cf. { Figures \ref{obsmod} and \ref{resid}}); this is because the density parameter is lower than in model 1 (Table \ref{modpar1}), for the reason discussed above. Model 3, plotted in Figure \ref{modobsApr16} in blue, reproduces the radio observations reasonably well at low frequencies, but not at high frequencies, particularly in the R polarization.

We note that all models failed to reproduce the peak observed $T_b$ of $4.7\times10^6$\,K. Seeking an explanation, we can propose additional, non-thermal emission. Indeed, there was a moving pore of opposite polarity at the NE of the large spot, marked by the arrow in Figure \ref{overview} (see also the HSP continuum image in Figure \ref{obsmod}, taken a few hours earlier); it is thus possible that the interaction of the moving pore with the big spot produced a small population of energetic particles, sufficient to give some weak non-thermal emission that made the brightness temperature approach 5\,MK, as in the case of the moving sunspot reported by \cite{1987SoPh..112...89C} (see also \citealp{2008SoPh..249..315U}). Additional support for this hypothesis is provided by the shift of the microwave source towards the location of the pore at long wavelengths.

The right panel of Figure \ref{modobsApr16} shows the variation of the physical parameters with height, up to the maximum height of the 3rd harmonic. In the same figure we have plotted in dashed lines the corresponding parameters from the sunspot model of \cite{2015ApJ...811...87A}. The differences in the low TR are due to the different values of the $H_0$ parameter. Higher up, model 1 gives higher temperatures than model 2, and model 3 still higher; this model was adjusted to reproduce the highest brightness temperature, for which it gives a value of $4\times10^6$\,K, higher than the prediction of the other models, yet lower than the observed. All models give temperatures well above the predictions of \cite{2015ApJ...811...87A}. Model 2 also gives a lower density than the others, but still above that of the \cite{2015ApJ...811...87A} model. The ordering of the pressure and the DEM is similar to that of the density, except that model 3 predicts an unphysical increase of the pressure with height and a decrease of the DEM; this increase is not inherent to the original \cite{2008A&A...488.1079S} model, but is a result of our modification of the temperature structure, in an effort to match the model results to the observations.

\begin{figure}
\begin{center}
\includegraphics[width=0.75\textwidth]{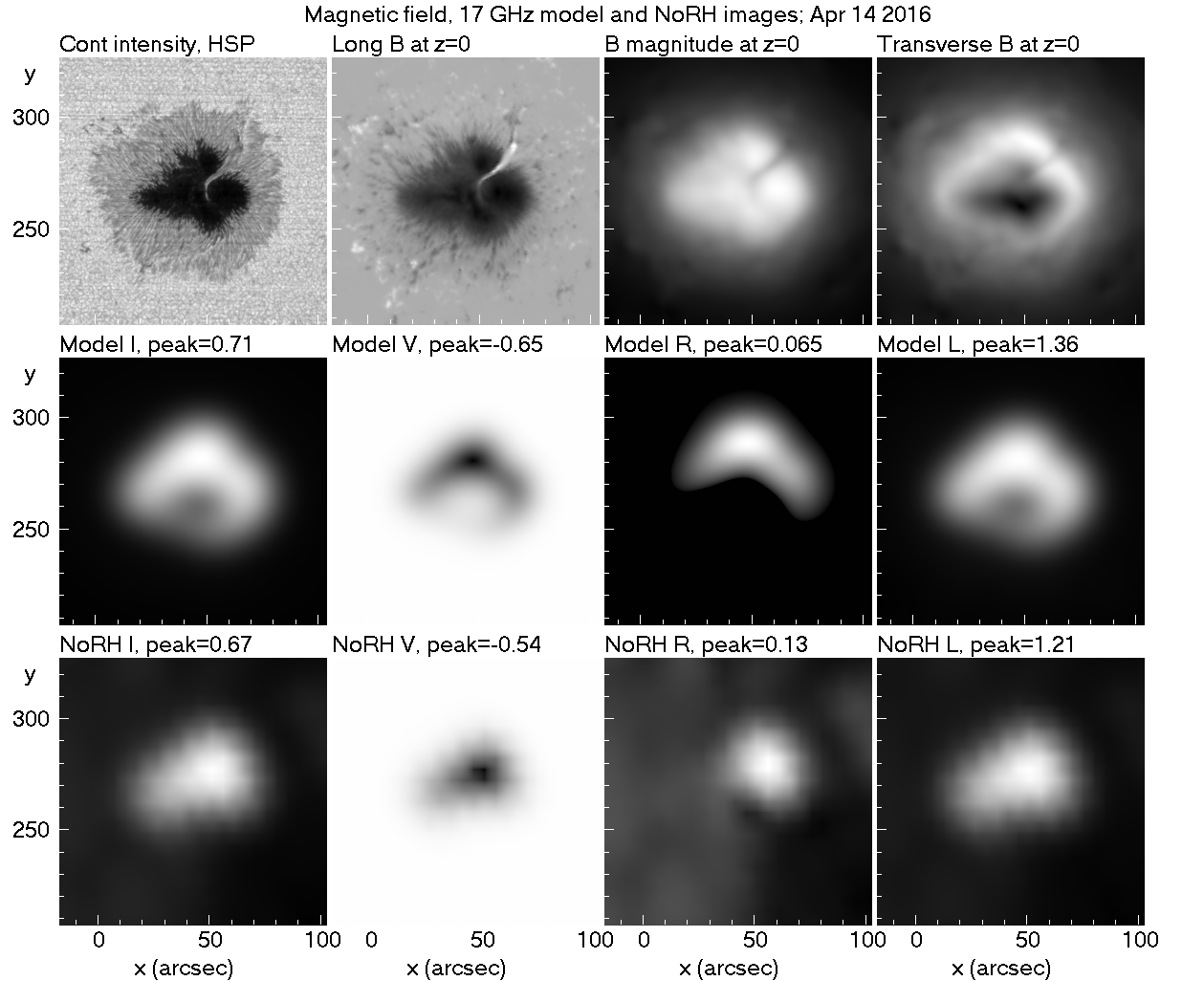}
\end{center}
\caption{Model of the NoRH observations. Top row: Hinode SP continuum intensity, longitudinal magnetic field, magnetic field intensity and transverse field at zero height. Middle row: computed 17\,GHz intensity for I, V, R and L, convolved with the NoRH beam. Bottom row: observed intensity, corrected for solar rotation. Model and observed NoRH images are displayed with the same contrast, with peak values in 10$^6$\,K printed in the image titles. Image orientation is with solar north up.}
\label{modelNoRH}
\end{figure}

\subsubsection{Modeling of the NoRH observations}
We used the parameters of the best $F_c$ model to compute model images at 17\,GHz, for comparison with the NoRH observations. Figure \ref{modelNoRH} shows the results, together with Hinode SP and NoRH images. Although the peak brightness temperature of the model and observed images are similar, the source shape is quite different. The model images are very similar to the transverse magnetic field image, proving the importance of the field orientation in the value of the third harmonic absorption coefficient, while the observations do not show such a similarity. These differences could not be remedied by changing the model parameters or by using a linear force free field extrapolation rather than a potential one. Images produced with Model 2 are morphologically  the same, with $T_b$ values smaller by a factor of $\sim2.5$ on the average.

We can offer two interpretations for the difference between model and observations. One is that the SW part of the model L source is invisible, due to the density being lower than the prediction of our plane-parallel model; in this case, the observed source would be displaced by 5\arcsec\ W and 4\arcsec\ S of the model source; this is small, compared to the NoRH resolution ($\sim$15\arcsec) but, in addition to the shift, the observed source is considerably broader than the model source. The other is that part of the observed emission could be associated to the opposite polarity light bridge, which is just below the L polarization source. As already noted in Section \ref{Model_RATAN}, we had intense activity at this location and this might have produced a small population of non-thermal emission, sufficient to produce some excess brightness. Alternatively, the magnetic field might be sheared to an extent that could not be accounted for by linear force-free extrapolations; this could change significantly its orientation and hence the gr emission. 

\section{Summary and conclusions}\label{concl}
In this article we presented a method that we developed to invert the differential emission measure obtained from AIA observations in the EUV and to compute the temperature and density structure along the line of sight, under the assumptions of stratification and hydrostatic equilibrium. For all three spots we studied, the DEM showed a strong peak near $2\times10^6$\,K, a weaker one near $4.5\times10^5$\,K and a still weaker near $9.5\times10^6$\,K which we ignored as unphysical. The peaks are at higher temperatures than in the sunspot DEM published by \cite{2009A&A...505..307T} from spectral data. The inversion reproduces well the sharp $T_e$ rise in the low TR and the slow rise higher up. 

We used the results of the DEM inversion, averaged over the sunspots and assumed a plane-parallel atmosphere to compute the microwave emission, which we then compared with RATAN-600 observations. For the magnetic field we used potential extrapolations of the longitudinal photospheric from the Hinode HSP and from HMI, with the method of \cite{1981A&A...100..197A}. We also modeled the microwave data with the constant conductive flux model of \cite{1980AA....82...30A} and { a variant of the active region} model of \cite{2008A&A...488.1079S} { and compared the 2D emission computed with the constant $F_c$ model for the sunspot of April 2016 with the NoRH image at 17\,GHz}.

The DEM model reproduced very well the observations of the moderate-size spot on October 2011 and within 25\% the data of a similar sized spot on March 2016, but gave too low antenna temperature and flux values for the big spot of April 14, 2016. 
On the other hand the  constant $F_c$ model, with inferred conductive flux values consistent with active region values from the \cite{1977ARA&A..15..363W} compilation, reproduced reasonably well the observed bulk parameters of the radio emission for all three sunspots that we tried; however, it could not reproduce the peak brightness temperature of $\sim5\times10^6$\,K observed in April 14, 2016 spot near 5\,GHz and the shape of the NoRH images of the same spot. We consider this as evidence for weak non-thermal emission associated with a moving pore of opposite polarity NW of the big sunspot and, possibly, with an opposite polarity light bridge that intruded well within the sunspot umbra. We further note that the DEM-based model is much less flexible than the constant $F_c$ model, which has three free parameters: still the DEM model is the only one of the two that incorporates $T_e$ and $N_e$ information from the EUV.

Our models were successful in reproducing the double-peak structure of the April 2016 sunspot, observed in the RATAN-600 1D spectra at high frequencies. We conclude that this structure is due to opacity changes, as the angle between the magnetic field and the line of sight changes over the sunspot, rather than due to variations of either the electron temperature or the electron density. No significant temperature variations over the sunspots were detected from our DEM analysis, while the average density above the umbra was only 1.3-1.5 times smaller than the average density over the sunspot, a value consistent with the results of \cite{2000ApJS..130..485N}. Of course, the presence of sunspot plumes at several locations near the umbra-penumbra boundary changes this picture, however density variations should not affect much the gyroresonance emission, once a particular harmonic layer is optically thick.

This work potentially opens new roads in the modeling of sunspot atmospheres in the transition region and low corona, which will be particularly useful as new, high resolution spectral imaging instruments such as the Expanded Owens Valley Solar Array, the extended VLA, the Siberian Radioheliograph (SRH) and the Chinese Mingantu Spectral Radioheliograph (MUSER) start providing data. On the EUV side, the DEM computation may be improved by using spectral data, i.e. from EIS. Finally, on the computational side, the plane-parallel assumption could be replaced by computing the atmospheric parameters at each point over the field of view; however, this may encounter problems due to the stratification assumption and a way must be devised to extend the model below the minimum temperature accessible through the AIA spectral bands.

\begin{acks}
The authors gratefully acknowledge use of data from the NoRH, Hinode and SDO (AIA and HMI) databases. C.E.A wishes to thank the St. Petersburg branch of Special Astrophysical Observatory (Spb SAO) and the Saint-Petersburg National Research University (ITMO) for their invitation and warm hospitality during his stay in Pulkovo. This work was also supported by the Russian state contract No AAAA-A17-117011810013-4 and the Russian Foundation for Basic Research grant No. 18-02-00045 (V.M.B., T.I.K.).
\end{acks}

\medskip\noindent{\footnotesize {\bf Disclosure of Potential Conflicts of Interest} The authors declare that they have no conflicts of interest.}

\end{article} 
\end{document}